\documentclass[twocolumn,pra,twocolumn,superscriptaddress]{revtex4}
\usepackage[latin9]{inputenc}
\usepackage{color}
\usepackage{amsmath}
\usepackage{amssymb}
\usepackage{bm}
\usepackage{braket}
\usepackage{graphicx}
\usepackage{tikz}
\usepackage{mathrsfs} 
\usepackage{float}
\usepackage{bbm}
\usepackage{epstopdf}
\usepackage{epsfig}
\usepackage{verbatim}
\usepackage{array}
\usepackage{setspace}
\usepackage{times}
\usepackage{algorithm}
\usepackage{algpseudocode}
\usepackage{booktabs}
\usepackage{multirow}
\usepackage{array}

\usepackage[unicode=true,
 bookmarks=true,bookmarksopen=false,
 breaklinks=false,pdfborder={0 0 1},backref=false,colorlinks=true]
 {hyperref}
\hypersetup{
 linkcolor=magenta, urlcolor=blue, citecolor=blue, pdfstartview={FitH}, hyperfootnotes=false, unicode=true}

\newcolumntype{C}[1]{>{\centering\arraybackslash$}p{#1}<{$}}

\begin{document}

\title{Accelerating Noisy Variational Quantum Algorithms with Physics-Informed Denoising Networks}

\author{Jie Liu}
\affiliation{Department of Physics, City University of Hong Kong, Tat Chee Avenue, Kowloon, Hong Kong SAR, China}
\affiliation{City University of Hong Kong Shenzhen Research Institute, Shenzhen, Guangdong 518057, China}
\affiliation{Quantum Science Center of Guangdong-Hong Kong-Macao Greater Bay Area, Shenzhen, Guangdong 518045, China}
\author{Xin Wang}
\email{x.wang@cityu.edu.hk}
\affiliation{Department of Physics, City University of Hong Kong, Tat Chee Avenue, Kowloon, Hong Kong SAR, China}
\affiliation{City University of Hong Kong Shenzhen Research Institute, Shenzhen, Guangdong 518057, China}

\date{\today}

\date{\today}

\begin{abstract}
Variational quantum algorithms are promising for near-term quantum computing, but are severely limited by hardware noise and the substantial circuit overhead required for error mitigation methods such as Zero-Noise Extrapolation (ZNE). We propose a Physics-Informed Denoising Network (PIDN) that reduces the cost of ZNE by learning a surrogate model of its optimization dynamics. By viewing the variational update as a trajectory in the parameter space, PIDN is trained to reproduce ZNE-mitigated expectation values and gradient directions while incorporating a physics-informed loss that preserves the gradient descent dynamics. Once trained, PIDN replaces repeated multi-noise evaluations with denoised expectation and gradient estimation directly from the current noisy observation and the historical trajectory, significantly reducing circuit executions. We benchmark the approach on the quantum approximate optimization algorithm for 3-regular graphs, Sherrington-Kirkpatrick, and transverse-field Ising models, as well as the variational quantum eigensolver for LiH, BeH$_2$ and H$_2$O. Across all tasks, PIDN attains performance comparable to ZNE, while reducing the number of circuit executions by a factor of approximately 4 to 6. Gradient cosine similarity with ZNE remains above 0.95 throughout training. Robustness analysis shows that PIDN fails only when ZNE itself becomes unreliable, and ablation studies confirm the necessity of the physics-informed loss for maintaining directional consistency. We further find that PIDN tracks optimization dynamics most accurately when the effective loss landscape retains strong low-frequency structure. These results establish PIDN as a scalable, resource-efficient strategy for noise-resilient variational optimization in the noisy intermediate-scale quantum regime.

\end{abstract}

\maketitle

\section{INTRODUCTION} \label{Intro}
Variational Quantum Algorithms (VQAs) have emerged as a leading paradigm for exploiting near-term quantum hardware to tackle optimization and simulation problems that are beyond the reach of classical computation \cite{cerezo2021variational,bauer2020quantum,bharti2022noisy,mcclean2016theory,melnikov2023quantum}. Operating within a hybrid quantum-classical framework, VQAs such as the Variational Quantum Eigensolver (VQE) for quantum chemistry and the Quantum Approximate Optimization Algorithm (QAOA) for combinatorial optimization employ Parameterized Quantum Circuits (PQCs) whose parameters are iteratively optimized by a classical optimizer to minimize a cost function $C(\bm{\theta})$, typically the expectation value $\langle H(\bm{\theta}) \rangle = \langle \psi(\bm{\theta})|H|\psi(\bm{\theta})\rangle$, of a problem Hamiltonian or objective observable $H$ \cite{peruzzo2014variational,farhi2014quantum,tilly2022variational,abbas2023quantum}. This flexible structure has enabled wide applications, positioning VQAs as flagship candidates for demonstrating quantum advantage on noisy intermediate-scale quantum (NISQ) devices \cite{preskill2018quantum,benamer2025variational}.

Despite their theoretical flexibility and inherent resilience to certain systematic biases, the practical deployment of VQAs on contemporary hardware faces a dual challenge that threatens their scalability: the high cost of quantum measurement and the fragility of the optimization landscape under hardware noise \cite{larocca2025barren,fontana2021evaluating,wang2021noise}. The primary obstacle to efficient VQA convergence is the corruption of the gradient signal. In practice, the cost function $C(\bm{\theta})$ is not measured exactly but is estimated through a finite number of measurement outcomes, or ``shots". This introduces stochastic shot noise, which scales as $1/\sqrt{N_\mathrm{shots}}$. To achieve the precision required for actual applications, the sampling requirements become prohibitively expensive. 

Furthermore, hardware-intrinsic noise, including gate errors, $T_1$/$T_2$ decoherence, and crosstalk, shifts the global minimum and flattens the landscape, often leading to the well-known phenomenon of barren plateaus \cite{knill2005quantum,wang2021noise,schumann2024emergence}. In the presence of such noise, the standard approach to gradient estimation, the parameter-shift rule, becomes highly unreliable. The parameter-shift rule requires evaluating the cost function at shifted coordinates, e.g., $\theta \pm \pi/2$. Since each evaluation is independently perturbed by both shot noise and hardware bias, the resulting gradient vector $\nabla C(\bm{\theta})$ often points in a suboptimal or even divergent direction.

Both shot and hardware noise sources can in principle be mitigated. Shot noise can be mitigated by increasing the number of measurements and hardware noise can be mitigated through error mitigation techniques such as Zero-Noise Extrapolation (ZNE). However, these remedies significantly inflate the ``shot tax" on quantum hardware \cite{giurgica2020digital,he2020zero}, namely the additional sampling overhead required to obtain reliable error-mitigated estimates. In particular, methods such as ZNE require the same observable to be measured repeatedly under multiple amplified noise settings, after which the results are combined through extrapolation. Consequently, the total number of circuit repetitions can become significantly larger than that required for a single unmitigated evaluation.

\begin{figure*}
	\includegraphics[width=\textwidth]{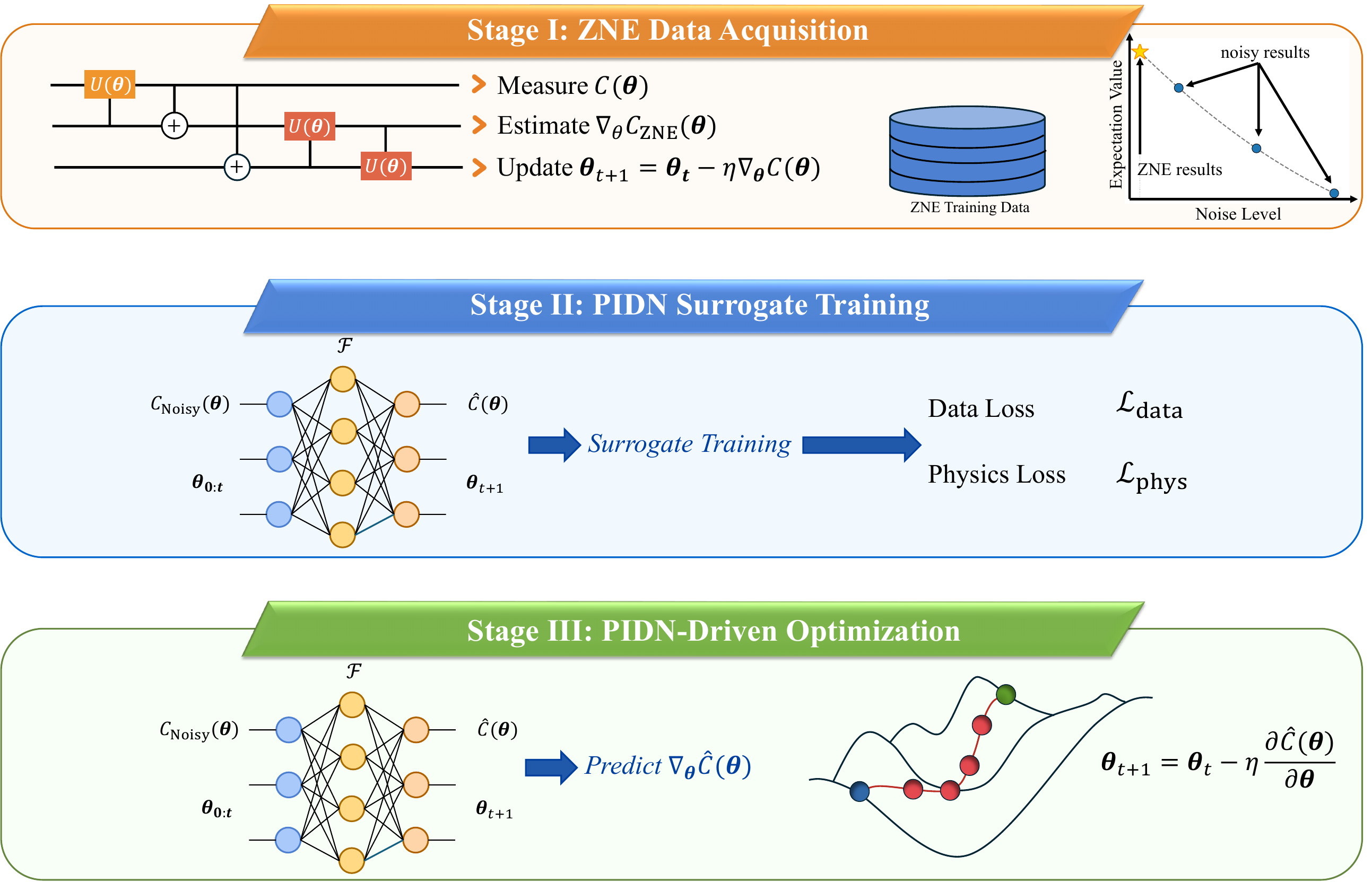}
	\caption{Unified three-stage workflow for physics-informed reconstruction of ZNE optimization.
Stage I (ZNE Data Acquisition) performs noise-scaled circuit evaluations to evaluate the zero-noise-extrapolated cost $C_\mathrm{ZNE}(\bm{\theta})$ and its gradient $\nabla_{\bm{\theta}}C_\mathrm{ZNE}(\bm{\theta})$, generating training data along the optimization trajectory. Stage II (PIDN Surrogate Training) uses these samples to train a neural surrogate $\mathcal{F}$ by minimizing a composite objective consisting of value-matching loss and gradient-consistency loss, thereby reconstructing the effective ZNE-induced optimization vector field. Stage III (PIDN-Driven Optimization) replaces further ZNE evaluations with classical predictions of surrogate gradients, enabling continued optimization with substantially reduced quantum resource consumption while preserving the geometric structure of the ZNE-corrected landscape.}
	\label{fig:illustration}
\end{figure*}

A key observation underpinning this work is that VQA optimization is inherently a dynamical process: parameters evolve along a correlated trajectory determined by the cost landscape, optimizer, and noise characteristics. Standard training protocols, however, treat each iteration independently, repeatedly estimating noisy expectation values and gradients from scratch \cite{sweke2020stochastic}. This ignores the strong temporal structure present in the optimization dynamics and leads to redundant measurements. Exploiting this structure offers an opportunity to reduce measurement overhead without compromising convergence.
Motivated by this insight, we propose a Physics-Informed Denoising Network (PIDN) that simultaneously denoises low-shot expectation values and predicts the next-step parameters along the VQA optimization trajectory. Rather than reducing iteration counts, our objective is to minimize the number of quantum measurements required per training step. The PIDN takes as input the previous circuit parameters $\bm{\theta}_0, \bm{\theta}_1,...,\bm{\theta}_t$ and the current noisy, low-shot expectation value $C_\mathrm{noisy}(\bm{\theta}_t)$, and outputs (i) a denoised estimate $\hat{C}(\bm{\theta}_t)$ and (ii) the predicted next-step parameters $\bm{\theta}_{t+1}$. Crucially, once trained, the model enables the optimizer to advance along the trajectory without performing explicit parameter-shift gradient evaluations at every step, relying instead on learned dynamics. The unified workflow for the PIDN is illustrated in Fig.~\ref{fig:illustration}. It is worth emphasizing that PIDN does not perform autonomous future prediction in the conventional time-series sense. At each iteration, the current noisy expectation value is still obtained from the quantum circuit and supplied to the model. The task is therefore more accurately understood as reconstructing the underlying ideal optimization trajectory conditioned on the current noisy observation, rather than extrapolating the future path solely from past data. This approach is closely related to surrogate modeling, where a computationally tractable classical model, often a neural network, is used to approximate an otherwise computationally expensive process, such as gradient descent with ZNE evaluations in our example. Surrogate models are widely employed to alleviate the burden of repeatedly evaluating costly objective functions, enabling faster optimization and more efficient exploration of the parameter space \cite{hirashima2023surrogate,Maganasurrogate}. By replacing expensive evaluations with a learned approximation, the iterative procedure is significantly accelerated while retaining the essential structure of the underlying problem.

To ensure reliability and physical consistency, the PIDN is trained using a hybrid loss function that combines two complementary objectives. First, a data-driven denoising loss supervises the network using high-fidelity reference values obtained from ZNE, teaching the model to reconstruct accurate expectation values from low-shot measurements. Second, a physics-informed loss enforces consistency with gradient-based optimization dynamics, constraining the predicted parameter updates to align with the expected descent direction dictated by $\nabla_{\bm{\theta}}\langle C(\bm{\theta})\rangle$. By embedding the structure of VQA optimization directly into the learning process, the PIDN can maintain stable training trajectories while predicting reliable optimization directions. 

This approach enables a different training paradigm: costly explicit quantum evaluations are used only in the first few iterations, while future steps are carried out using the learned surrogate dynamics with low cost quantum evaluations. As a result, the total shot cost scales mostly with the number of anchoring evaluations rather than the number of parameters or training steps, yielding substantial measurement shot savings.

Our contributions are threefold: First, we propose a physics-informed neural network architecture tailored to VQA training that integrates measurement denoising with parameter prediction in a unified framework. Second, we formulate a hybrid loss function that balances data fidelity (via ZNE-supervised denoising) with physical consistency (via gradient descent constraints), enabling the model to act as a reliable dynamics surrogate. Third, we validate our approach through extensive experiments on benchmark problems, including QAOA for different combinatorial problems and VQE for different molecules, demonstrating 80\% reductions in the number of shots compared to standard training protocols while achieving comparable or superior convergence accuracy. 

The remainder of this paper is organized as follows: Sec.~\ref{sec:preliminaries} provides background on VQAs, noise sources, and physics-informed machine learning; Sec.~\ref{sec:methods} details our proposed PINN architecture and loss formulation; Sec.~\ref{sec:results} presents results; and we conclude in Sec.~\ref{sec:conclusion}.

\section{Preliminaries}\label{sec:preliminaries}
\subsection{Variational Quantum Algorithms and Optimization Dynamics} VQAs solve optimization and simulation problems by minimizing a parameterized quantum cost function within a hybrid quantum-classical loop. Given a PQC $U(\bm{\theta})$, a parameterized quantum state is prepared \cite{benedetti2019parameterized}:
\begin{equation}
	|\psi(\bm{\theta})\rangle=U(\bm{\theta})|0\rangle^{\otimes n},
\end{equation}
and the cost function is defined as the expectation value of a Hermitian observable $H$:
\begin{equation}
	\langle H(\bm{\theta}) \rangle = \langle \psi(\bm{\theta})|H|\psi(\bm{\theta})\rangle.
\end{equation}
In practice, $H$ is often expressed as a weighted sum of Pauli operators \cite{reggio2024fast},
\begin{equation}
	H = \sum_{k=1}^{M} h_k P_k,
\end{equation}
so that the cost function $C(\bm{\theta})$ can be decomposed as:
\begin{equation}
	C(\bm{\theta}) = \sum_{k=1}^{M} h_k \langle P_k(\bm{\theta}) \rangle. \label{eq:expansion}
\end{equation}
We use QAOA and VQE as two benchmarks in this study.

In QAOA, the cost Hamiltonian encodes a classical optimization problem or spin model. Below we specify three representative Hamiltonians considered in this work.
\paragraph{MaxCut on 3-Regular Graphs.}

For a graph $G = (V, E)$ (where $V$ and $E$ denote the sets of vertices and edges, respectively), the MaxCut Hamiltonian is
\begin{equation}
    H_{\mathrm{MaxCut}}
    =
    \sum_{(i,j)\in E}
    \frac{1}{2}(I - Z_i Z_j),
\end{equation}
where each vertex is associated with a qubit, and $Z_i Z_j$ denotes the tensor-product Pauli-$Z$ operator acting on the qubits corresponding to vertices $i$ and $j$.
Maximizing $\langle H_{\mathrm{MaxCut}} \rangle$ corresponds to maximizing the number of edges crossing the cut \cite{guerreschi2019qaoa}.

The performance of QAOA is quantified using the Approximation Ratio (AR):
\begin{equation}
    \mathrm{AR}
    =
    \frac{
    \langle H_{\mathrm{MaxCut}} \rangle
    }{
    C_{\max}
    },
    \label{eq:AR1}
\end{equation}
where $C_{\max}$ is the optimal classical cut value. $\mathrm{AR}\in[0,1]$ with value close to $1$ indicating near-optimal performance.

\paragraph{Sherrington--Kirkpatrick (SK) Model.}
The SK model is an infinite-range spin glass defined as \cite{aizenman2003extended}
\begin{equation}
    H_{\mathrm{SK}}
    =
    \sum_{i<j}
    J_{ij} Z_i Z_j,
\end{equation}
where $i$ and $j$ are spin indices, and $Z_i Z_j$ denotes the tensor-product Pauli-$Z$ operator acting on spins $i$ and $j$. Here, $J_{ij}$ is the coupling strength between the $i$-th and $j$-th spins, randomly drawn from a normal distribution:
\begin{equation}
    J_{ij} \sim \mathcal{N}(0,1).
\end{equation}

The objective is to minimize $\langle H_{\mathrm{SK}} \rangle$, and performance is again measured via AR to the ground state energy:
\begin{equation}
    \mathrm{AR}
    =
    \frac{E_{\mathrm{gs}}}{\langle H_{\mathrm{SK}} \rangle}
    ,
    \label{eq:AR2}
\end{equation}
where $E_{\mathrm{gs}}$ is obtained via classical exact diagonalization for small systems. $\mathrm{AR}\in[0,1)$ with value close to $1$ indicating near-optimal performance.

\paragraph{Transverse Field Ising Model (TFIM).}
The one-dimensional TFIM is defined as \cite{arezzo2026digital}:
\begin{equation}
    H_{\mathrm{TFIM}}
    =
    -J \sum_{i=1}^{n-1}
    Z_i Z_{i+1}
    -
    h \sum_{i=1}^{n}
    X_i,
\end{equation}
where $i$ is the spin index, $Z_i Z_{i+1}$ denotes the product of Pauli-$Z$ operators acting on neighboring spins $i$ and $i+1$, $X_i$ is the Pauli-$X$ operator acting on spin $i$, $J$ is the interaction strength, and $h$ is the transverse field strength.

For TFIM, the objective is to minimize the energy $\langle H_{\mathrm{TFIM}} \rangle$. Performance is evaluated using the approximation ratio relative to the exact ground state energy:
\begin{equation}
    \mathrm{AR}
    =
    \frac{
    E_{\mathrm{gs}}}{\langle H_{\mathrm{TFIM}} \rangle},
    \label{eq:AR3}
\end{equation}
$\mathrm{AR}\in[0,1)$ with value close to $1$ indicating near-optimal performance.

In this work, we use the quantum alternating operator ansatz for the QAOA problems, which uses a $p$-layer circuit with two alternating operators as PQCs.

In the VQE, the Hamiltonian corresponds to a molecular electronic structure problem \cite{loaiza2023reducing}:
\begin{equation}
    H_\mathrm{molecule}
    =
    \sum_{pq}
    h_{pq}
    a_p^\dagger a_q
    +
    \frac{1}{2}
    \sum_{pqrs}
    h_{pqrs}
    a_p^\dagger a_q^\dagger a_r a_s,
\end{equation}
where $p,q,r,s$ index the spin-orbitals of a chosen single-particle basis. The coefficients $h_{pq}$ and $h_{pqrs}$ denote the one-electron and two-electron integrals, corresponding to the kinetic energy and electron-nuclear interaction terms, and the electron-electron Coulomb interaction, respectively. The operators $a_p^{\dag}$ and $a_p$ are fermionic creation and annihilation operators that create and remove an electron in spin-orbital $p$. After fermion-to-qubit transformation (e.g., Jordan-Wigner) \cite{tranter2018comparison}, this Hamiltonian is mapped to a Pauli decomposition:
\begin{equation}
    H_\mathrm{molecule}
    =
    \sum_{\ell=1}^{L}
    w_\ell P_\ell.
\end{equation}

For example, the LiH molecule in a minimal basis may require 4 qubits and hundreds of Pauli terms \cite{shee2022qubit}.

The VQE objective is to minimize the energy:
\begin{equation}
    E(\bm{\theta})
    =
    \langle \psi(\bm{\theta}) |
    H
    | \psi(\bm{\theta}) \rangle.
\end{equation}

The performance of VQE is evaluated by the energy error:
\begin{equation}
    \Delta E
    =
    E(\bm{\theta})
    -
    E_{\mathrm{FCI}},
    \label{eq:energy_err}
\end{equation}
where $E_{\mathrm{FCI}}$ is the exact Full Configuration Interaction energy \cite{knowles1984new}.

A solution is said to achieve \emph{chemical accuracy} if
\begin{equation}
    \Delta E
    \leq
    1.6 \times 10^{-3}\ \mathrm{ Hartree},
\end{equation}
corresponding to approximately $1$ kcal/mol, which is sufficient for predicting reaction energetics in quantum chemistry \cite{dalton2024quantifying}.

In this work, the variational ansatz used for molecular simulations is the Unitary Coupled Cluster with Singles and Doubles (UCCSD) ansatz, which is widely adopted in quantum chemistry due to its systematic improvability and chemical interpretability \cite{grimsley2019trotterized,fan2023circuit}.

VQA training proceeds iteratively by updating the parameters according to a classical optimization rule. For gradient descent, this takes the form
\begin{equation}
	\bm{\theta}_{t+1} = \bm{\theta}_{t}-\eta \nabla_{\bm{\theta}} C(\bm{\theta}),
\end{equation}
where $\eta$ is the learning rate and $t$ indexes the optimization step. This recursion defines a discrete-time dynamical system in parameter space, with the optimization trajectory $\{\bm{\theta}\}_{t=0}^{T}$ determined by both the cost landscape and the chosen optimizer.

Gradients are commonly evaluated using the parameter-shift rule, which allows exact differentiation of expectation values for gates of the form $e^{-i\theta_i G_i}$ with $G_i^2=I$:
\begin{equation}
	\frac{\partial C(\bm{\theta})}{\partial \theta_i} = \frac{1}{2}\bigg( 
	\left\langle H(\bm{\theta}+\frac{\pi}{2}\bm{e}_i) \right \rangle-\left\langle H(\bm{\theta}-\frac{\pi}{2}\bm{e}_i) \right \rangle \bigg),
\end{equation}
Consequently, evaluating the full gradient requires $2n$ additional circuit executions to estimate the expectation value per optimization step for an ansatz with $n$ parameters. Each of these evaluations must itself be estimated from repeated quantum measurements, making the overall training cost scale unfavorably with circuit size, number of parameters and training duration.
\subsection{Shot Noise, Hardware Noise, and Zero-Noise Extrapolation}
Expectation values on quantum hardware are estimated from a finite number of measurement shots $N_\mathrm{shots}$. For a single observable $P_k$ with outcomes $m_j \in \{\pm 1\}$, the estimator
\begin{equation}
	\hat{P}_k = \frac{1}{N}\sum_j^{N} m_j
\end{equation}
has variance
\begin{equation}
	\mathrm{Var}(\hat{P}_k) = \frac{1-\langle P_k\rangle^2}{N},
\end{equation}
implying that the total variance of the cost function estimator scales as \cite{kandala2017hardware,wecker2015progress}:
\begin{equation}
	\mathrm{Var}(\hat{C}) = \sum_k h_k^2 \mathrm{Var}(\hat{P}_k)\sim\mathcal{O}\bigg(\frac{1}{N}\bigg).
\end{equation}
Reducing shot noise thus requires increasing the number of measurements 
$N$, which directly increases experimental cost. It is also worth mentioning that the variance of the cost function grows with the number of terms in the expansion of the Hamiltonian $H$ as shown in Eq.~\eqref{eq:expansion}, making accurate evaluation of larger systems even more demanding \cite{gonthier2022measurements}.

In addition to statistical fluctuations, hardware noise introduces systematic errors through decoherence, control imperfections, and crosstalk. Let $\mathcal{E}_\lambda$ denote a noisy quantum channel parameterized by an effective noise strength $\lambda$. The measured expectation value can be written as
\begin{equation}
	C_\lambda(\bm{\theta})=\mathrm{Tr}(H\mathcal{E}_\lambda(\rho(\bm{\theta}))),
\end{equation}
which deviates from the ideal value $C_0(\bm{\theta})$.

ZNE mitigates this bias by deliberately amplifying noise and extrapolating to the zero-noise limit. In a typical ZNE workflow \cite{temme2017error,majumdar2023best,anand2023classical,pascuzzi2022computationally}:
\begin{enumerate}
	\item The circuit is executed at multiple noise amplification factors $\{\lambda_1,\lambda_2,...,\lambda_L\}$, often implemented via gate folding or pulse stretching \cite{li2017efficient,temme2017error}.
	\item Expectation values $C_{\lambda_l}(\bm{\theta})$ are measured at each noise level.
	\item A model (e.g., linear or polynomial) is fit to the data and extrapolated to $\lambda \to 0$,
	\begin{equation}
		C_\mathrm{ZNE}(\bm{\theta}) = \lim_{\lambda\to 0} C_{\lambda}(\bm{\theta}).
	\end{equation}
	\end{enumerate}
While ZNE can significantly reduce systematic bias, it increases the total shot cost by a factor proportional to the number of noise levels $L$ and must be repeated at each optimization step. When combined with gradient-based training, the overall measurement overhead becomes prohibitive for practical VQA implementations.

Recent advances in quantum error mitigation have increasingly explored deep-learning-based methods, and an important distinction among these methods lies in the assumptions they make about how noise depends on the hardware, the circuit structure, and the variational parameters \cite{zlokapa2020deep,liao2024machine,liao2025sample}. These assumptions directly determine how the training dataset is generated and, consequently, what aspect of the noisy quantum process the learned model is expected to capture. One broad class of approaches assumes that the dominant source of noise is primarily hardware dependent. In this case, the goal is to learn a correction model that reflects the response of a specific device, rather than the behavior of a particular ansatz \cite{zlokapa2020deep,liao2024machine}. The dataset is therefore constructed by running many circuits on the same hardware to obtain noisy outputs, while classically simulating the same circuits to obtain the corresponding ideal outputs, or using traditional quantum error mitigation methods such as ZNE to get less noisy outputs if the underlying circuits are hard to classically simulate. These noisy-clean pairs are then used to train a neural network that takes as input features such as the circuit structure, the circuit parameters, and the associated noisy measurement results, and outputs a corrected estimate. Because the model is trained to capture the characteristic noise behavior of the hardware itself, this type of method is relatively general and can, in principle, be applied to many different circuits executed on the same device.

A second class of methods assumes that the noise depends not only on the hardware, but also strongly on the circuit structure. Under this assumption, the training data must preserve structural similarity to the target circuit, since the learned model is intended to capture how noise acts on circuits of a particular form. One important example is the family of Clifford-based methods, inspired by Clifford data regression \cite{Czarnik2021errormitigation,angusdCDR}. The central idea is that circuits with similar structures are expected to be affected by noise in similar ways, even if their detailed gate content is different. Based on this idea, one generates many classically simulable Clifford circuits that mimic the structure of the target circuit, executes them on the noisy hardware to obtain noisy outputs, and pairs these with their ideal outputs obtained from classical simulation. The resulting training set allows the model to learn a structure-dependent correction rule without requiring exact classical simulation of the original non-Clifford target circuit. A related but distinct situation arises in fixed-structure VQAs, where the ansatz architecture is held fixed and the circuit is evaluated over many different parameter settings. In this case, the noisy and ideal data pairs are generated by sweeping across parameter space for the same variational circuit structure \cite{liao2025sample}. The objective here is typically to train a surrogate model that approximates the parameter-to-observable map, or even a broad region of the variational loss surface itself \cite{shaffer2023surrogate, liao2025sample, du2025efficient,liao2025demonstration}. Because the model must represent the behavior of the VQA over an extended region of parameter space, this type of approach generally requires training data that span a sufficiently broad portion of the landscape, and its data cost can therefore grow rapidly with the number of variational parameters.

A third class of methods adopts a trajectory-based viewpoint. Rather than attempting to learn a correction model that is valid for arbitrary circuits on a device, or a surrogate that reconstructs the full loss surface of a fixed ansatz, trajectory-based methods restrict attention to the region of parameter space that is actually visited during optimization \cite{huang2025palqo}. Accordingly, the dataset is generated by collecting samples only along the optimization trajectory itself. In geometric terms, one may think of this as learning not the entire surface, but only a narrow ``trajectory tube'' surrounding the dynamically relevant path on that surface. This significantly reduces the amount of training data required, because the model is no longer asked to reconstruct the full global landscape. Instead, it only needs to capture the local optimization behavior relevant to the variational dynamics. This strategy is particularly attractive when the ultimate goal is not to predict arbitrary expectation values everywhere in parameter space, but to accelerate the optimization process itself. However, the trajectory-based approach in Ref.~\cite{huang2025palqo} focuses only on mitigating shot noise and does not address hardware-induced noise. In this sense, existing deep-learning-based mitigation methods differ not only in what they aim to correct, but also in the assumptions they make, the way their datasets are constructed, and the aspect of the noisy quantum process that the model is trained to learn. In contrast, our method treats noisy VQA training itself as a dynamical system and uses physics-informed learning to reconstruct both denoised expectation values and the associated parameter evolution under the combined effect of shot noise and hardware noise.

These three classes of learning-based mitigation methods reflect different tradeoffs between generality, structural specificity, and data efficiency. Hardware-dependent approaches aim for broad applicability across circuits executed on the same device, but may require large and diverse training datasets in order to capture the device response over many circuit families. Structure-dependent approaches exploit the similarity between circuits of a given form, which can improve correction accuracy for a target ansatz or circuit class, but at the cost of reduced transferability beyond that structure. Trajectory-based approaches are the most targeted in their data usage, since they restrict learning to the dynamically relevant region of parameter space explored during optimization, thereby avoiding the need to reconstruct the full variational landscape. In this sense, the choice of method depends strongly on the ultimate objective: whether one seeks hardware-level generality, structure-aware correction, or data-efficient acceleration of a specific variational optimization process.

In this work, we build on the trajectory-based perspective and further tailor it to the setting of noisy variational quantum algorithms under realistic hardware error sources. Our method treats VQA training itself as a dynamical system and uses physics-informed learning to jointly reconstruct denoised expectation values and the associated parameter evolution. In this way, the learned model is constrained not only to correct the observed cost values, but also to preserve the optimization dynamics induced by noise-mitigated training. This perspective moves beyond point-wise correction of isolated observables and instead learns the structured optimization flow in the dynamically relevant region of parameter space. As a result, our framework captures both local information in the energy landscape and the trajectory-level behavior of the variational update process, enabling efficient emulation of ZNE-assisted optimization with substantially reduced circuit executions.

\subsection{Physics-Informed Neural Network} 
Physics-Informed Neural Networks (PINNs) incorporate known physical laws into neural network training by enforcing them as soft constraints in the loss function \cite{raissi2019physics,lu2021deepxde,de2022error,karniadakis2021physics}. Given a differential equation of the form
\begin{equation}
	\mathcal{D}[u(\bm{x},t)]=0,
\end{equation}
a PINN approximates the solution $u_{\bm{w}}(\bm{x},t)$ with a neural network parameterized by $\bm{w}$ and minimizes a composite loss
\begin{equation}
	\mathcal{L} = \mathcal{L}_{\mathrm{data}}+\lambda \mathcal{L}_{\mathrm{physics}}, \mathcal{L}_{\mathrm{physics}}=\bigg|\bigg|\mathcal{D}[u_{\bm{w}}(\bm{x},t)]\bigg|\bigg|^2,  
\end{equation}
which adds a penalty to the network when $u_{\bm{w}}(\bm{x},t)$ violates the differential equation while minimizing the discrepancy $\mathcal{L}_\mathrm{data}$ between the collected data and the prediction.

Beyond solving static differential equations, PINNs have been successfully applied to learning dynamical systems and time-stepping operators, where the network predicts the next-step solution $u(\bm{x}, t+\Delta t)$ given the current state $u(\bm{x}, t)$, while enforcing the underlying evolution law \cite{cai2021physics,nath2023physics}. This formulation has enabled PINN-based models to predict system dynamics, infer unknown parameters, and reconstruct governing behavior even when the available data are sparse or noisy \cite{nath2023physics,chen2021physics,liu2025hamiltonian}.

\begin{figure*}
	\includegraphics[width=\textwidth]{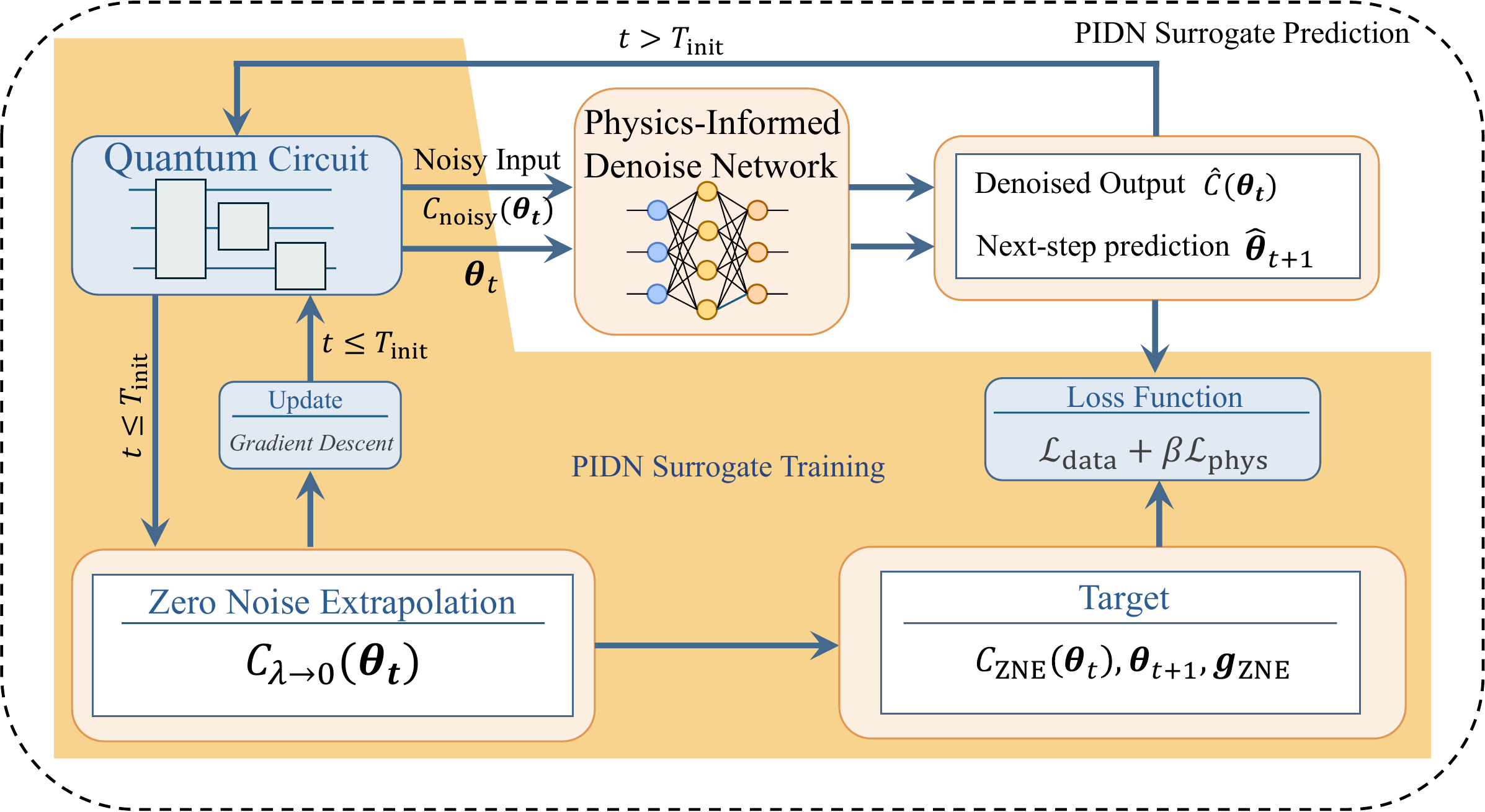}
	\caption{Workflow and architecture of the physics-informed denoising network (PIDN).
Schematic illustration of the PIDN-assisted optimization framework. At early iterations ($t\leq T_{\mathrm{init}}$), ZNE is performed to generate denoised reference values and gradients, which serve as training targets. The PIDN takes two inputs: the noisy expectation value obtained from the quantum circuit and the historical parameter trajectory encoded via a GRU branch. These representations are fused to produce two outputs: a denoised expectation value approximating the ZNE-corrected cost and a prediction of the next-step parameters. The network is trained using a composite loss function combining supervised reconstruction and a physics-informed term. After the training phase ($t > T_{\mathrm{init}}$), PIDN replaces ZNE to provide denoised cost estimates and parameter updates, thereby reducing circuit-execution overhead while preserving optimization dynamics.}
	\label{fig:workflow}
\end{figure*}
\section{Methods}
\label{sec:methods}
In this section, we present a unified workflow of PIDN that integrates VQAs, ZNE, and PINN, as introduced in Sec.~\ref{sec:preliminaries}. The objective of PIDN is to reconstruct the effective noise-mitigated optimization dynamics using a learned surrogate model trained on partial ZNE data, thereby reducing quantum resource consumption while preserving optimization fidelity.

PIDN accelerate the training of noisy VQAs by learning an effective, noise-mitigated optimization flow in the parameter space. It is worth noting that this method aims not to accelerate convergence and reduce the number of training iterations, but rather aims to reduce the number of evaluations of the underlying circuit within each iteration and accelerate the experimental process and alleviate the burden of data collection to make meaningful updates to the trainable parameters utilizing fewer shots. At its core, the PIDN treats the parameter update process of a VQA as a discrete-time dynamical system subject to both stochastic shot noise and systematic hardware noise. Motivated by this, PIDN formulates the parameter update as next-step prediction of a dynamical system while trying to denoise the expectation value and minimize the effect of noise.

The PIDN is trained together during the VQA optimization.
In the early training iterations, ZNE is employed to provide high-fidelity reference data 
\begin{equation}
	\bigg(\bm{\theta}_t, C_\mathrm{noisy}(\bm{\theta}_t),C_{\mathrm{ZNE}}(\bm{\theta}_t),\bm{\theta}_{t+1}\bigg).
\end{equation}
ZNE is employed throughout the first $T_\mathrm{init}$ iterations, namely $t=0,1,2,...T_\mathrm{init}$. These reference data are then fed into the PIDN, which simultaneously performs (i) denoising of expectation values, i.e., $C_\mathrm{noisy}(\bm{\theta}_t) \rightarrow C_{\mathrm{ZNE}}(\bm{\theta}_t)$ and (ii) prediction of the next-step variational parameters, i.e., $\bm{\theta}_t \rightarrow \bm{\theta}_{t+1}$. Besides minimizing the discrepancy between the output value and the reference value provided by ZNE,
physics-informed constraints derived from gradient-based optimization are also incorporated to regularize the learned dynamics, the effect of which will be analyzed later. 

After training, the PIDN is used to predict future parameters update from the current noisy measurement, and then the parameters are fed into the circuit to provide the next noisy measurement under limited budget without employing ZNE. The PIDN and circuit are then alternately evaluated to optimize the parameters in the PQC until convergence. The workflow of PIDN is presented in Fig.~\ref{fig:workflow}.



\subsection{Problem Formulation}

Consider a parameterized quantum circuit
\begin{equation}
    |\psi(\bm{\theta})\rangle = U(\bm{\theta}) |0\rangle,
\end{equation}
where $\bm{\theta} \in \mathbb{R}^d$ denotes the variational parameters.

The ideal cost function is
\begin{equation}
    C(\bm{\theta}) 
    = \langle \psi(\bm{\theta}) | H | \psi(\bm{\theta}) \rangle,
\end{equation}
while under hardware noise the measured objective becomes $C_\mathrm{noisy}(\bm{\theta})$. Applying ZNE with noise scaling factors 
$\{\lambda_k\}_{k=1}^{L}$ produces an estimate of the noise-mitigated cost $C_{\mathrm{ZNE}}(\bm{\theta})$.
Define the corresponding gradients:
\begin{align}
    \bm{\mathrm{g}}_{\mathrm{Noisy}}(\bm{\theta})
    &= \nabla_{\bm{\theta}} C_\mathrm{noisy}(\bm{\theta}), \\
    \bm{\mathrm{g}}_{\mathrm{ZNE}}(\bm{\theta})
    &= \nabla_{\bm{\theta}} C_{\mathrm{ZNE}}(\bm{\theta}).
\end{align}

Our goal is to learn a surrogate model 
\begin{equation}
    \mathcal{F}_{\phi} : \bigg(\bm{\theta}_{0:t}, C_\mathrm{noisy}(\bm{\theta}_t)\bigg) \mapsto \bigg( \hat{C}(\bm{\theta}_t), \hat{\bm{\theta}}_{t+1} \bigg)
\end{equation}
parameterized by $\phi$, such that its induced gradients $\nabla_{\bm{\theta}_t} \hat{C}(\bm{\theta}_t)$
approximate $\bm{\mathrm{g}}_{\mathrm{ZNE}}(\bm{\theta})$ in the region explored during optimization and at the same time minimize the difference between $(\hat{C}(\bm{\theta}_t),\hat{\bm{\theta}}_{t+1})$ and $(C_{\mathrm{ZNE}}(\bm{\theta}_t),\bm{\theta}_{t+1})$ .

This surrogate aims to enable replacement of repeated ZNE evaluations by classical predictions in the future stage after sufficient training.

\subsection{Physics-Informed Surrogate Training}
First, we use ZNE to collect training data. During the initial training phase, ZNE is applied by evaluating the circuit at multiple noise levels $\lambda > 1$ and extrapolating to the zero-noise limit to get the trajectory $\bm{\theta}_{0:T_\mathrm{init}}$ under gradient descent, the noisy expectation value $C_\mathrm{noisy}(\bm{\theta}_{0:T_\mathrm{init}-1})$, the denoised expectation value $C_{\mathrm{ZNE}}(\bm{\theta}_{0:T_\mathrm{init}-1})$ and the corresponding gradients $\bm{\mathrm{g}}_{\mathrm{ZNE}}(\bm{\theta}_{0:T_\mathrm{init}-1})$.

Each optimization step using ZNE provides a supervised training pair
\begin{equation}
\left(
\underbrace{C_\mathrm{noisy}(\bm{\theta}_{0:t}), \bm{\theta}_{0:t}}_{\mathrm{input}},
\underbrace{C_{\mathrm{ZNE}}(\bm{\theta}_t), \bm{\theta}_{t+1}^{(0)},\bm{\mathrm{g}}_{\mathrm{ZNE}}(\bm{\theta}_t)}_{\mathrm{target}}
\right),
\end{equation}
where $\bm{\theta}_{t+1}^{(0)}$ denotes the next-step parameters obtained from a gradient descent using the extrapolated cost function.

Following the PINN framework described in Sec.~II.C, we construct a composite loss function that enforces both data fidelity and gradient consistency.

Given a dataset
\begin{equation}
\left\{
C_\mathrm{noisy}(\bm{\theta}_{0:t}), \bm{\theta}_{0:t},
C_{\mathrm{ZNE}}(\bm{\theta}_t), \bm{\theta}_{t+1},\bm{\mathrm{g}}_{\mathrm{ZNE}}(\bm{\theta}_t)
\right \}_{t=0}^{T_\mathrm{init}-1},
\end{equation}
which is collected from ZNE-corrected evaluations along an optimization trajectory, we define the supervised data loss
\begin{equation}
    \mathcal{L}_{\mathrm{data}}
    =
    \frac{1}{T_\mathrm{init}}
    \sum_{t=0}^{T_\mathrm{init}-1}
    \left\|
    \hat{C}(\bm{\theta}_t)-C_{\mathrm{ZNE}}(\bm{\theta}_t)\right\|^2+
    \left\| \hat{\bm{\theta}}_{t+1}
    - \bm{\theta}_{t+1} \right\|_2^2,
\end{equation}
and the physics-informed gradient consistency loss
\begin{equation}
    \mathcal{L}_{\mathrm{phys}}
    =
    \frac{1}{T_\mathrm{init}}
    \sum_{t=0}^{T_\mathrm{init}-1}
    \left\|
    \nabla_{\bm{\theta}_t} \hat{C}(\bm{\theta}_t)
    -
    \bm{\mathrm{g}}_{\mathrm{ZNE}}(\bm{\theta}_{t})
    \right\|_2^2.
\end{equation}
Here, $\|\cdot\|_2^2$ denotes the squared Euclidean norm, i.e., the sum of the squared components of the vector.
The total training objective is
\begin{equation}
    \mathcal{L}(\phi)
    =
    \mathcal{L}_{\mathrm{data}}
    +
    \beta \,
    \mathcal{L}_{\mathrm{phys}},
\end{equation}
where $\beta$ controls the relative importance of enforcing gradient alignment.

This formulation is a direct specialization of the general PINN objective to our problem formulation setting, where gradient consistency enforces reconstruction of the effective noise-mitigated optimization flow.

The proposed PIDN adopts a dual-branch, dual-head architecture designed to jointly model observable correction and parameter dynamics. The input consists of two components: (i) a feedforward branch that takes the noisy expectation value as input, capturing instantaneous measurement information, and (ii) a recurrent branch implemented using a small GRU \cite{cho2014learning} that encodes the historical parameter trajectory $(\bm{\theta}_{0:t},C_\mathrm{noisy}(\bm{\theta}_{0:t}))$, thereby learning the temporal structure of the optimization process. These two latent representations are fused and mapped to two output heads: one predicts the denoised expectation value approximating the ZNE-mitigated cost, and the other predicts the next-step variational parameters, effectively learning the underlying optimization vector field. This joint prediction enforces consistency between cost reconstruction and parameter updates. The network is trained using the Adam optimizer to minimize a composite loss function combining supervised reconstruction and physics-informed constraints.

\subsection{Workflow}

The overall workflow consists of three stages:

\subsubsection{Stage I: ZNE Data Collection}

Starting from an initial parameter $\bm{\theta}_{0}$, we perform ZNE-corrected evaluations along the optimization trajectory. At iteration $t$:

\begin{enumerate}
    \item Evaluate $C_{\lambda_k}(\bm{\theta}_{t})$ for all noise levels $\{\lambda_k\}$.
    \item Compute the extrapolated cost $C_{\mathrm{ZNE}}(\bm{\theta}_{t})$.
    \item Estimate the gradient $\bm{\mathrm{g}}_{\mathrm{ZNE}}(\bm{\theta}_{t})$ using the parameter-shift rule applied to each noise-scaled circuit.
    \item Update parameters using gradient descent:
    \begin{equation}
        \bm{\theta}_{t+1}
        =
        \bm{\theta}_{t}
        -
        \eta \, \bm{\mathrm{g}}_{\mathrm{ZNE}}(\bm{\theta}_{t}).
    \end{equation}
\end{enumerate}

The resulting trajectory provides training samples for the surrogate model.

\subsubsection{Stage II: Surrogate Training}

Using dataset $\mathcal{D}$, we minimize $\mathcal{L}(\phi)$ to obtain trained parameters $\phi^{*}$.

\subsubsection{Stage III: Surrogate-Driven Optimization}

After training, PIDN is embedded into the workflow of VQA training, subsequent optimization steps replace ZNE evaluations by surrogate predictions:
\begin{equation}
    \bm{\theta}_{t+1}
    =
    \bm{\theta}_{t}
    -
    \eta \,
    \nabla_{\bm{\theta}} 
    \mathcal{F}_{\phi^{*}}(\bm{\theta}_{0:t},C_\mathrm{noisy}(\bm{\theta}_t)),
\end{equation}
and then $\bm{\theta}_{t+1}$ is fed into the ansatz to evaluate the noisy cost function $C(\bm{\theta}_{t+1})$. The ansatz and PIDN is alternately evaluated to optimize the circuit parameters.

This significantly reduces the number of shots, as $C_\mathrm{ZNE}$ is approximated by the surrogate model and gradient computations are performed classically.

\begin{figure*}[t]
	\includegraphics[width=\textwidth]{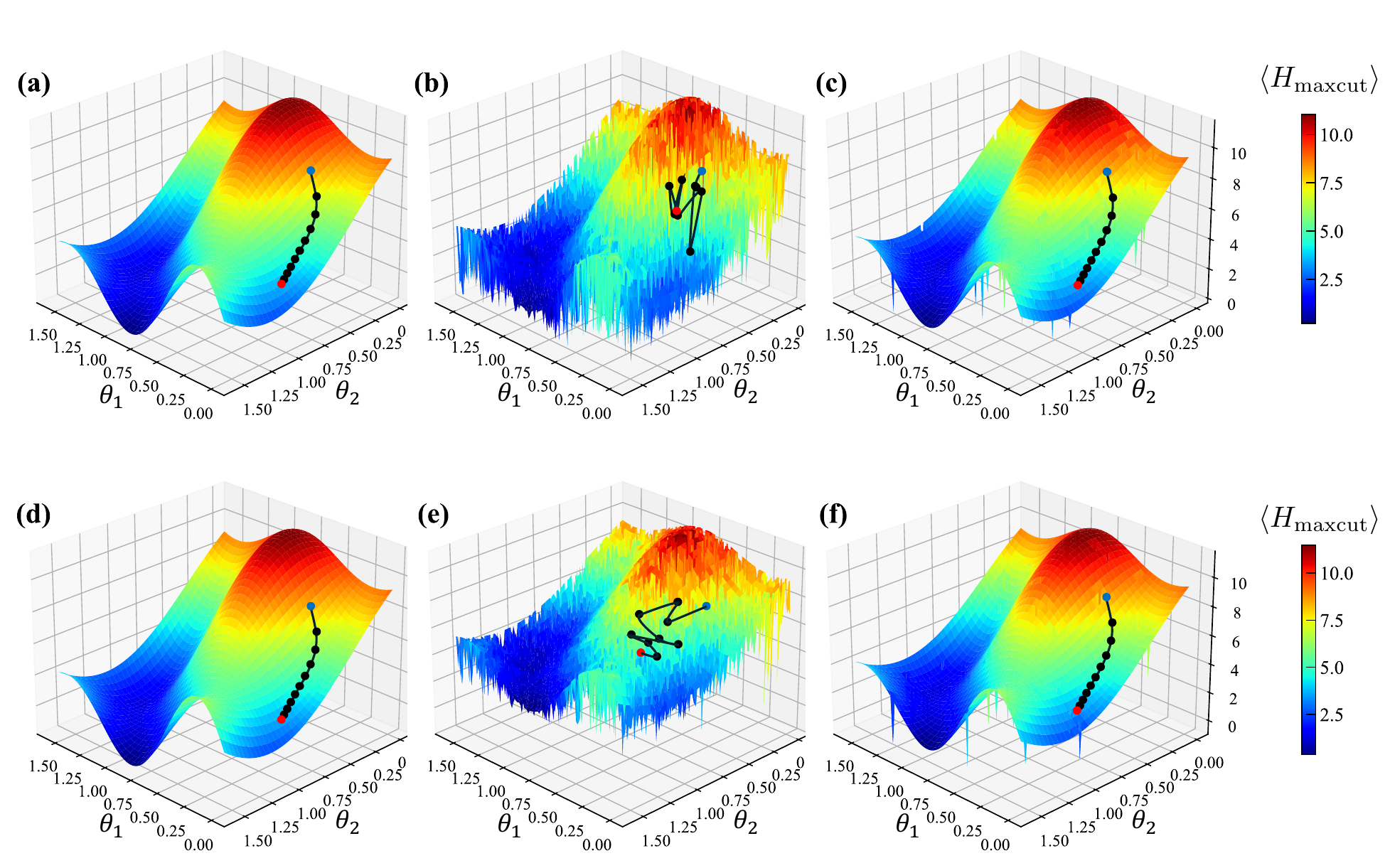}
	\caption{Noise-induced distortion of the optimization landscape.
(a-c) Loss landscapes and optimization trajectories  for (a) noiseless VQA, (b) noisy VQA, and (c) ZNE-assisted VQA under depolarizing noise. (d-f) Corresponding results under dephasing noise, arranged in the same order as (a-c). 
The optimization trajectories are shown as black and colored connected dots, with the blue dots being the starting points and red ones being ending points.
Noise deforms the cost landscape and alters optimization trajectories relative to the noiseless case, while ZNE partially restores the landscape geometry, yielding trajectories that more closely follow the noiseless dynamics. }
	\label{fig:landscape}
\end{figure*}

\section{RESULTS} \label{sec:results}
In this section, we benchmark the proposed PIDN across both combinatorial optimization and quantum chemistry tasks, focusing on its ability to reconstruct noise-mitigated optimization dynamics in ZNE-mitigated VQA and to reduce quantum measurement costs under realistic noise. We consider two representative classes of VQAs: the QAOA for combinatorial optimization problems and the VQE for molecular ground-state estimation. Throughout this section, performance is evaluated using task-appropriate metrics: for QAOA, we use AR, [see Eqs.~\eqref{eq:AR1}, \eqref{eq:AR2}, \eqref{eq:AR3}], with higher values indicating better performance; for VQE, we use the energy error [see Eq.~\eqref{eq:energy_err}], where lower values indicate better convergence.

All results are obtained under gate-level depolarization or dephasing noise, and comparisons are made between noisy VQA (denoted as Noisy in all figures below), ZNE-assisted VQA (denoted as ZNE in all figures below), and the proposed PIDN-assisted VQA (denoted as PIDN in all figures below) under consistent measurement budgets unless otherwise specified.
\subsection{Reconstruction of Effective Optimization Flow under Noise}\label{subsec:landscape}

We first investigate whether PIDN can faithfully reconstruct the effective optimization flow induced by ZNE, rather than merely reproducing final expectation values. This distinction is crucial: variational optimization depends not only on the loss landscape itself, but also on the local gradient directions that guide parameter updates.

To investigate how noise modifies the optimization geometry, we consider the MaxCut problem of 3-regular graphs solved using a QAOA ansatz. To enable visualization, we randomly fix all variational parameters except two ($\theta_1$ and $\theta_2$) and evaluate the objective function $\langle H_{\mathrm{MaxCut}}\rangle$ over a two-dimensional grid in this reduced parameter subspace. We compare three settings: Ideal (noiseless) VQA shown in Fig.~\ref{fig:landscape}(a) and (d), noisy VQA under depolarizing and dephasing noise shown in Fig.~\ref{fig:landscape}(b) and (e) and ZNE-assisted VQA under depolarizing and dephasing noise shown in Fig.~\ref{fig:landscape}(c) and (f).

In the noiseless case, the cost landscape exhibits smooth curvature and coherent basin structure. The surface varies continuously across the parameter domain, with well-defined gradients directing the optimization trajectory toward a local minimum. The curvature is structured and globally informative, indicating a well-conditioned optimization geometry.

In contrast, the noisy landscape exhibits two distinct geometric distortions:
\begin{itemize}
	\item Global flattening: Across large regions of parameter space, the cost values concentrate within a narrow range, significantly reducing contrast between high- and low-energy regions. This suppresses large-scale curvature and diminishes the directional information encoded in the landscape.
	\item Local roughness: At smaller scales, the surface becomes irregular and highly fluctuating. The cost function displays sharp, localized variations that are not present in the ideal landscape, resulting in high-frequency perturbations superimposed on an otherwise flattened structure.
\end{itemize}

Formally, this corresponds to a reduction in global gradient norm combined with increased variance in local gradient direction. As a result, the optimization geometry becomes ill-conditioned: the large-scale structure is suppressed, while small-scale noise-induced perturbations dominate the gradient signal.

When ZNE is applied, we observe a partial restoration of the global landscape structure. The basin geometry and large-scale curvature become visibly closer to the noiseless case. The overall energy contrast across parameter space increases relative to the noisy case, indicating recovery of meaningful global features.

However, residual artifacts remain: small regions exhibit sharp localized dips or irregular curvature not present in the ideal landscape. These features likely originate from extrapolation instability and amplified statistical fluctuations at higher noise scaling factors. Thus, while ZNE restores global geometry, it does not fully eliminate all distortions.

To systematically compare the loss landscapes in the ideal, ZNE-assisted and noisy VQA, we introduce quantitative metrics that separately capture global structure and local irregularity under dephasing noise. Let $E(\boldsymbol{\theta})$ denote the objective function $\langle H_{\mathrm{MaxCut}}\rangle$ evaluated on a discrete parameter grid $\boldsymbol{\theta} = (\theta_1,\theta_2)$.

\paragraph{Global structure (smoothed contrast).}
To isolate large-scale geometric features, we consider a Gaussian-smoothed surface
\begin{equation}
\tilde{E}(\boldsymbol{\theta}) = \mathcal{G}_{\sigma} * E(\boldsymbol{\theta}),
\end{equation}
where $\mathcal{G}_{\sigma}$ denotes a Gaussian kernel with width $\sigma$. We then define:
\begin{align}
\Delta E_{\mathrm{smooth}} &= \max_{\boldsymbol{\theta}} \tilde{E}(\boldsymbol{\theta}) - \min_{\boldsymbol{\theta}} \tilde{E}(\boldsymbol{\theta}), \\
\sigma_{\mathrm{smooth}} &= \sqrt{\mathrm{Var}\big[\tilde{E}(\boldsymbol{\theta})\big]}.
\end{align}
These quantities measure the effective energy contrast and variability of the large-scale landscape, suppressing high-frequency noise. A reduction in these metrics indicates global flattening.

\paragraph{Spectral structure.}
To quantify the distribution of spatial frequencies, we compute the two-dimensional Fourier transform $\hat{E}(\mathbf{k})$ and define the low-frequency power ratio:
\begin{equation}
R_{\mathrm{LF}} = \frac{\sum_{\mathbf{k} \in \mathcal{K}_{\mathrm{low}}} |\hat{E}(\mathbf{k})|^2}{\sum_{\mathbf{k}} |\hat{E}(\mathbf{k})|^2},\label{eq:RLF}
\end{equation}
where $\mathcal{K}_{\mathrm{low}}$ denotes a region near the origin in frequency space. This metric quantifies the fraction of energy contained in large-scale (low-frequency) structures. A decrease in $R_{\mathrm{LF}}$ indicates a transfer of energy toward high-frequency components, corresponding to increased local irregularity.

\begin{table}[b]
\centering

\begin{tabular}{c c c c}
\hline
VQA & $\Delta E_{\mathrm{smooth}}$ & $\sigma_{\mathrm{smooth}}$ & $R_{\mathrm{LF}}$ \\
\hline
Ideal       & 11.15 & 3.093 & 0.9798 \\
noisy    & 10.28 & 2.768 & 0.8021 \\
ZNE-assisted   & 11.12 & 3.086 & 0.9758 \\

\hline
\end{tabular}
\caption{Quantitative metrics of the loss landscape for three types of VQAs under depolarization noise.}
\label{tab:landscape_metrics}
\end{table}

\begin{table}[b]
\centering

\begin{tabular}{c c c c}
\hline
VQA & $\Delta E_{\mathrm{smooth}}$ & $\sigma_{\mathrm{smooth}}$ & $R_{\mathrm{LF}}$ \\
\hline
Ideal       & 11.15 & 3.093 & 0.9798 \\
noisy    & 10.23 & 2.752 & 0.7885 \\
ZNE-assisted   & 10.91 & 3.064 & 0.9757 \\

\hline
\end{tabular}
\caption{Quantitative metrics of the loss landscape for three types of VQAs under dephasing noise.}
\label{tab:landscape_metrics_dephasing}
\end{table}

Table~\ref{tab:landscape_metrics} and Table~\ref{tab:landscape_metrics_dephasing} summarize the quantitative landscape metrics for the three VQA settings. The ideal case is identical in both tables and serves as the common noiseless reference. In this limit, the landscape exhibits large values of $\Delta E_{\mathrm{smooth}}$ and $\sigma_{\mathrm{smooth}}$, together with a high low-frequency power ratio $R_{\mathrm{LF}} \approx 0.98$, indicating that the loss landscape is dominated by coherent large-scale structure and possesses a well-defined global basin geometry.

In the noisy VQA setting, all three metrics are significantly reduced for both depolarization and dephasing noise. Under depolarization noise, $\Delta E_{\mathrm{smooth}}$ decreases from $11.15$ to $10.28$, $\sigma_{\mathrm{smooth}}$ decreases from $3.093$ to $2.768$, and $R_{\mathrm{LF}}$ drops from $0.9798$ to $0.8021$. A similar but slightly stronger degradation is observed under dephasing noise, where $\Delta E_{\mathrm{smooth}}$ decreases to $10.23$, $\sigma_{\mathrm{smooth}}$ decreases to $2.752$, and $R_{\mathrm{LF}}$ is further reduced to $0.7885$. These changes indicate suppression of the global energy contrast, reduced large-scale variation, and a substantial redistribution of spectral weight away from low-frequency components. Taken together, these results show that noise degrades the global organization of the landscape and enhances local irregularity.

By contrast, the ZNE-assisted results remain close to the ideal reference in both noise models. For depolarization noise, the recovered values $\Delta E_{\mathrm{smooth}} = 11.12$, $\sigma_{\mathrm{smooth}} = 3.086$, and $R_{\mathrm{LF}} = 0.9758$ are nearly indistinguishable from the ideal case. Under dephasing noise, ZNE likewise restores the landscape metrics to near-ideal values, with only slightly larger residual deviations, particularly in $\Delta E_{\mathrm{smooth}}$. This demonstrates that ZNE effectively recovers the coherent low-frequency structure of the loss landscape and largely restores its global basin geometry.

Overall, the two tables show a consistent picture: noise induces a structural degradation of the optimization landscape rather than a simple uniform rescaling, while ZNE substantially reverses this effect. The primary impact of noise is to weaken large-scale basin features and increase the relative contribution of higher-frequency fluctuations, thereby making the landscape less favorable for efficient variational optimization. ZNE mitigates this distortion and preserves the large-scale geometric structure that is important for successful optimization.

To further analyze the effect of geometric distortion on optimization dynamics, we perform gradient descent from identical initial parameters in all three settings and plot the resulting trajectories over the corresponding landscapes, shown as black and colored connected dots, with the blue dots being the starting points and red ones being ending points.

In the noiseless case, the trajectory follows a smooth, monotonic descent path. The updates are coherent and progressively approach a local minimum, consistent with a stable and informative gradient field.

For ZNE-assisted optimization, the trajectory closely resembles the noiseless case. The descent direction remains stable across iterations, and the path converges toward a similar local basin. Minor deviations occur due to residual local roughness, but the overall flow structure is preserved.

In contrast, the noisy optimization trajectory exhibits qualitatively different behavior. Successive updates show inconsistent directional movement, with the parameter vector changing direction unpredictably between iterations. Rather than following a coherent descent path, the trajectory appears to wander irregularly across the surface.

This behavior can be understood geometrically: due to global flattening, the gradient magnitude is reduced across most regions, while local roughness introduces stochastic directional components. It can also be observed that the parameter update in the noisy VQA is much greater than those in both noiseless VQA and ZNE-assisted VQA, and must take a much smaller learning rate to make meaningful updates. This provides further evidence that local roughness leads to a larger gradient norm. Consequently, the effective gradient vector at each point is dominated by local fluctuations rather than global basin structure. Optimization thus resembles a random walk over a nearly flat landscape, eventually stagnating in a region where gradients become negligible.

To evaluate whether the proposed PIDN surrogate can faithfully reconstruct the ZNE-induced optimization dynamics, we perform the following protocol.

We first execute ZNE-assisted gradient descent for 45 iterations and collect the corresponding dataset
\begin{equation}
\left\{
C_\mathrm{noisy}(\bm{\theta}_t), \bm{\theta}_{0:t},
C_{\mathrm{ZNE}}(\bm{\theta}_t), \bm{\theta}_{t+1},\bm{\mathrm{g}}_{\mathrm{ZNE}}(\bm{\theta}_t)
\right \}_{t=0}^{45}.
\end{equation}
This dataset is used to train the PIDN surrogate model as described in Sec.~\ref{sec:methods}.

After training, subsequent parameter updates are performed using surrogate gradients $    \nabla_{\bm{\theta}} \mathcal{F}_{\phi}(\bm{\theta},C_\mathrm{noisy}(\bm{\theta}))$, thereby replacing further ZNE evaluations.

To quantify how well PIDN reconstructs the ZNE optimization flow, we compute the cosine similarity between gradients predicted by PIDN and gradients obtained from an independent ZNE optimization run:

\begin{equation}
	\mathrm{CosSim}(\bm{\mathrm{g}}_\mathrm{PIDN},\bm{\mathrm{g}}_\mathrm{ZNE}) = \frac{\bm{\mathrm{g}}_\mathrm{PIDN} \cdot \bm{\mathrm{g}}_\mathrm{ZNE}}{\|\bm{\mathrm{g}}_\mathrm{PIDN}||\times||\bm{\mathrm{g}}_\mathrm{ZNE}\|}.
	\label{eq:cossim}
\end{equation}

\begin{figure}
\includegraphics[width=\linewidth]{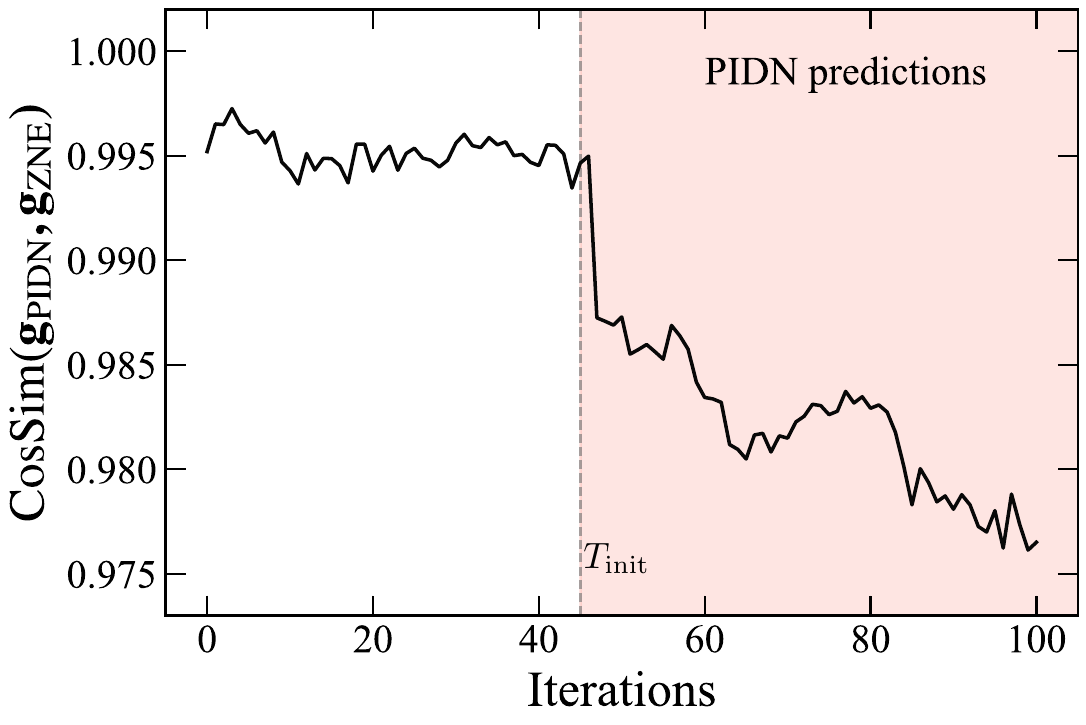}
\caption{Cosine similarity between PIDN-predicted gradients and ZNE gradients as a function of the number of iterations.}
\label{fig:cossim}
\end{figure}
Importantly, the ZNE gradients used for comparison are generated from the $45$th iteration and are not part of the training dataset, ensuring that the evaluation probes generalization rather than memorization. The result is shown in Fig.~\ref{fig:cossim}.

We observe that during the initial 45 iterations, when updates are still performed using ZNE, the cosine similarity between the PIDN-predicted gradients and ZNE gradients remains extremely high, approximately 0.995. This indicates that the surrogate has successfully learned the local structure of the ZNE-induced vector field within the explored region of parameter space.

After iteration 45, when PIDN takes over as the optimization driver, the cosine similarity gradually decreases but remains above 0.975 throughout subsequent iterations. We also observe that both the gradient magnitudes and denoised expectation values predicted by PIDN closely match those obtained from ZNE throughout the optimization process. In other words, PIDN does not merely approximate cost values; it approximates the ZNE-induced parameter update dynamics. This supports the interpretation that PIDN functions as a physics-informed reconstruction of the effective optimization flow, enabling continuation of ZNE-like training with significantly reduced quantum evaluations.


\subsection{Shot and Evaluation Efficiency Across VQA Tasks}
\begin{figure}
\includegraphics[width=\linewidth]{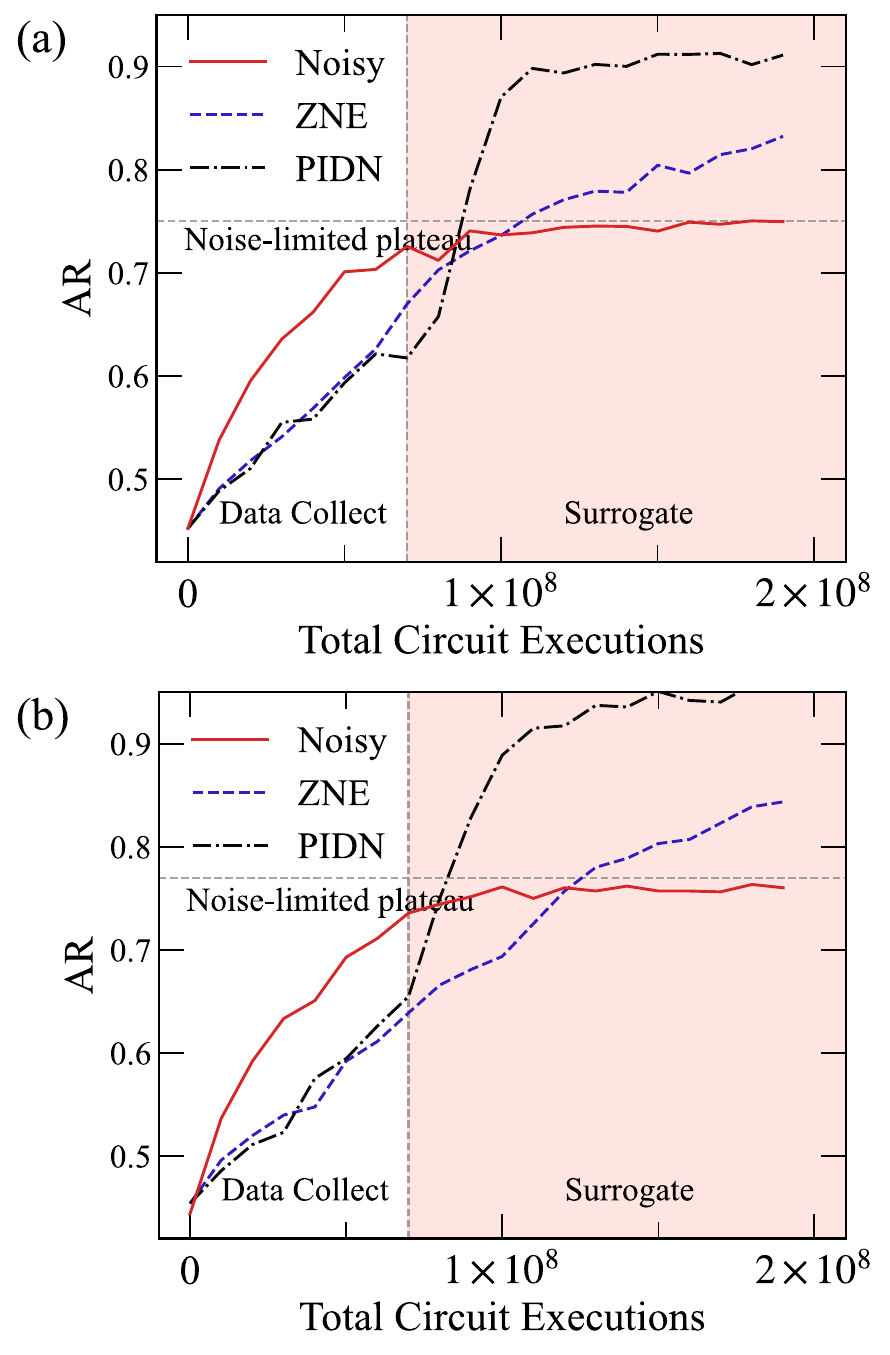}
	\caption{AR as a function of total circuit executions for QAOA on 3-regular graph instances for noisy-VQA, ZNE-assisted VQA and PIDN-assisted VQA under depolarizing noise with strength $1\times10^{-3}$ in (a) and under dephasing noise with strength $1\times10^{-3}$ in (b).
}
	\label{fig:QAOA_comp}
\end{figure}
In this subsection, we quantify the execution efficiency of PIDN-assisted VQA and compare it against both noisy VQA and ZNE-assisted VQA. The central question is not only whether PIDN reproduces ZNE-like performance, but whether it does so with substantially fewer circuit executions.

We first consider QAOA applied to 3-regular graph instances. Performance is evaluated using the AR as a function of the total number of circuit executions as shown in Fig.~\ref{fig:QAOA_comp}(a) under depolarization noise and Fig.~\ref{fig:QAOA_comp}(b) under dephasing noise. The AR of noisy VQA initially increases most rapidly as a function of total executions. This behavior is expected: noisy VQA does not allocate additional circuit executions to noise extrapolation. All executions are used directly for parameter updates. Consequently, under a fixed execution budget, noisy VQA performs more optimization steps. However, this rapid growth quickly saturates. The AR plateaus at a relatively low value and shows no significant improvement even as the number of circuit execution. This behavior is consistent with the landscape analysis in Sec.~\ref{subsec:landscape}: under depolarizing noise, the cost landscape becomes globally flattened and locally irregular. The directions of gradients are divergent at local irregularities, leading to premature convergence around $\mathrm{AR}\approx 0.75$ under depolarization noise and around $\mathrm{AR}\approx 0.78$ under dephasing noise.

In contrast, ZNE-assisted VQA exhibits a slower initial increase in AR. This is due to the overhead of evaluating the circuit at multiple noise scaling factors in order to estimate the zero-noise expectation and its gradient. For the same total number of executions, fewer parameter updates are performed compared to noisy VQA.

Nevertheless, ZNE achieves a significantly higher final AR. The reconstructed gradient field restores the global curvature of the loss landscape, enabling continued descent beyond the noisy plateau.

PIDN-assisted VQA exhibits a two-phase behavior: 
\begin{itemize}
	\item Phase I (Data Collection Phase): During the initial training period, PIDN relies on ZNE-generated data. Consequently, its AR growth closely follows that of ZNE.
	\item Phase II (Surrogate Phase): After PIDN becomes the effective update model, the growth rate of AR increases dramatically. Since PIDN directly predicts the zero-noise update direction without requiring multi-noise evaluations, the execution overhead disappears. Each circuit evaluation contributes directly to optimization progress.
\end{itemize}
It is worth noting that the AR growth rate in this phase becomes even faster than the early stage of noisy VQA, because the update direction retains ZNE-like global structure, but without the execution overhead. PIDN rapidly converges to $\mathrm{AR}\approx 0.9$, matching the final saturated performance of ZNE, while using substantially fewer total executions.

\begin{table}[b]
\centering
\begin{tabular}{ccc}
\hline
Speedup & Average  & Std.  \\
\hline
TFIM (QAOA)            & 5.8              & 0.7                 \\

SK-model (QAOA)       & 5.3              & 0.5                \\

3-regular (QAOA)      & 5.6              & 0.6                 \\

LiH (VQE)			& 4.4				& 0.9
\\

BeH$_2$ (VQE) &5.2 &1.1
\\

H$_2$O (VQE) &5.6 &1.4
\\
\hline
\end{tabular}
\caption{Impact of PIDN on QAOA and VQE speedup.
Average and standard deviation of speedup for QAOA and VQE of PIDN relative to ZNE on different problem instances under depolarization and dephasing noise.}
\label{tab:qaoaspeedup}
\end{table}

To quantify efficiency more quantitatively, we measure the total number of circuit executions required for ZNE and PIDN to reach a target $\mathrm{AR}=0.9$. This analysis is performed across three models: 3-regular graphs, SK model and  TFIM.

For multiple instances and parameter initializations, we compute the speedup:
\begin{equation}
	\mathrm{Speedup}
	=
\frac{\mathrm{No.\ of\ Executions\ required\ by\ ZNE}}
{\mathrm{No.\ of\ Executions\ required\ by\ PIDN}},
\end{equation}
and we report mean and standard deviation across instances in TABLE~\ref{tab:qaoaspeedup}, with the observed average speedups being $5.6$ for 3-regular graphs, $5.3$ for SK model and $5.8$ for TFIM. 

These results demonstrate that PIDN consistently reduces execution cost by approximately a factor of five across structurally different combinatorial optimization tasks.

We next conduct a similar study for VQE simulations of molecular ground-state energy estimation using the UCCSD ansatz under depolarizing noise.

\begin{figure}
	\includegraphics[width=\linewidth]{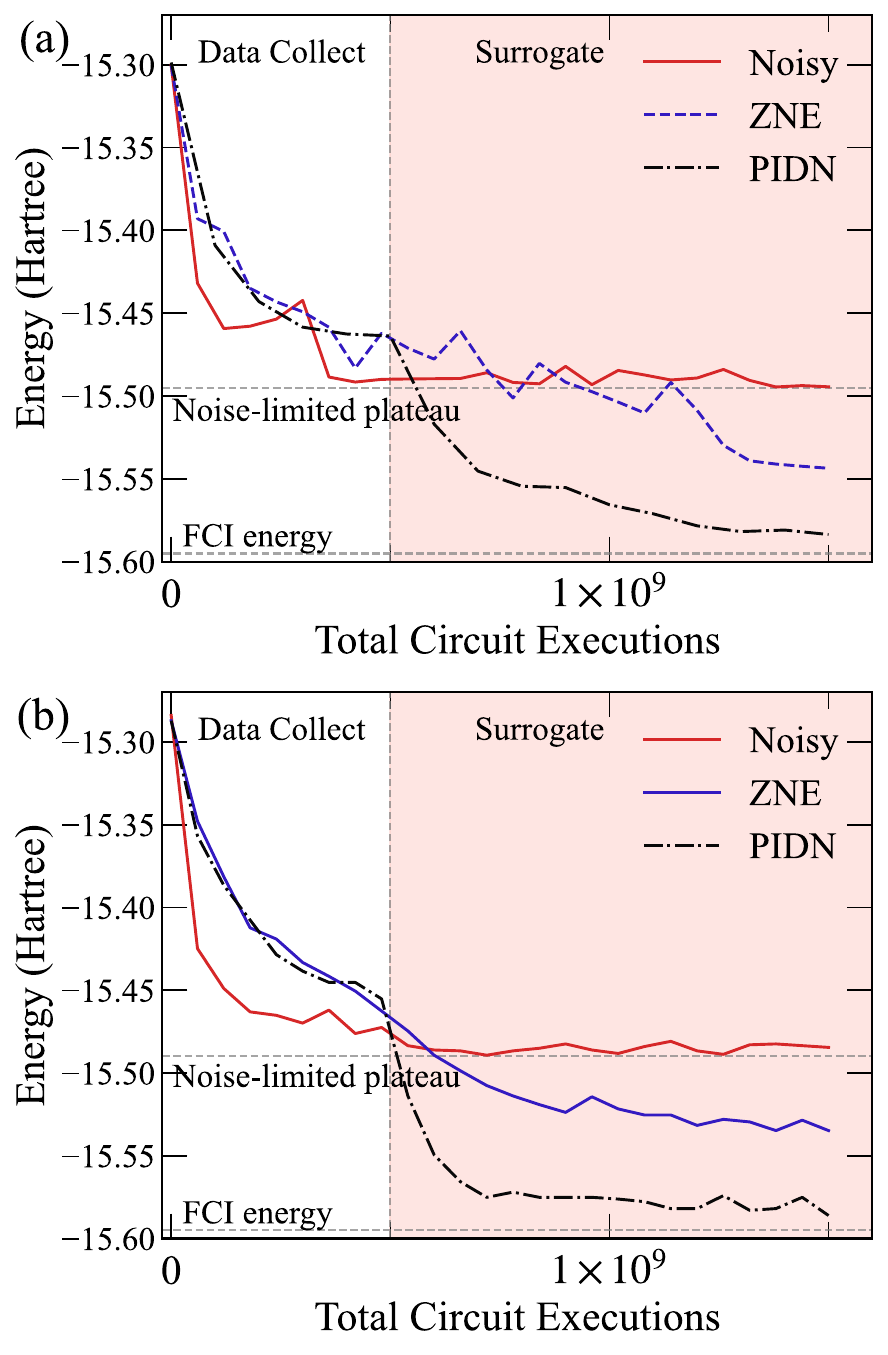}
	\caption{Ground-state energy convergence for BeH$_2$ as a function of total circuit executions for noisy-VQA, ZNE-assisted VQA and PIDN-assisted VQA under depolarization noise with strength $1\times10^{-3}$ in (a) and under dephasing noise with strength $1\times10^{-3}$ in (b)}
	\label{fig:VQE_comp}
\end{figure}

For VQE, the AR-like metric used in QAOA is replaced by the variational energy, measured in Hartree (Ha). The qualitative trends are similar to those observed for QAOA, as shown in Fig.~\ref{fig:VQE_comp}. Under both depolarization and dephasing noise, noisy VQE exhibits a rapid initial decrease in energy but saturates early, reaching only $E(\bm{\theta})\approx -15.49~\mathrm{Ha}$ in the depolarization case and $E(\bm{\theta})\approx -15.48~\mathrm{Ha}$ in the dephasing case. In contrast, ZNE-assisted optimization converges more gradually but attains substantially lower energies, approaching $-15.55~\mathrm{Ha}$ under both noise models. The PIDN-assisted approach initially follows a trajectory similar to that of ZNE and then accelerates after the transition to surrogate-based updates, ultimately reaching the lowest final energy of approximately $-15.58~\mathrm{Ha}$.

We measure the speedup required to reach a fixed energy accuracy threshold across three molecules as show in TABLE~\ref{tab:qaoaspeedup}, with values $4.4$ for $\mathrm{LiH}$, $5.2$ for BeH$_2$ and $5.6$ for H$_2$O.
Despite increases in ansatz complexity and system size, the observed speedup remains consistently in the range of $4.4-5.8$, suggesting that PIDN's efficiency gain scales favorably with problem complexity.

Overall, these results demonstrate that PIDN fundamentally reshapes the cost-accuracy tradeoff inherent to error-mitigated VQAs. While noisy optimization is execution-efficient but performance-limited, and ZNE restores performance at substantial evaluation overhead, PIDN achieves the best of both regimes: it preserves the optimization geometry reconstructed by ZNE while eliminating the repeated multi-noise extrapolation cost during later stages of training. The consistently observed $4\sim6$ fold reduction in circuit executions across combinatorial optimization problems and molecular VQE tasks indicates that the learned surrogate captures structure in the noise-induced deformation along the optimization flow. Importantly, the speedup remains stable across different problem classes and system sizes, suggesting that PIDN does not merely accelerate a specific instance but provides a scalable mechanism for reducing the cost of error mitigation. These findings establish PIDN-assisted optimization as a practically viable pathway toward resource-efficient, noise-resilient variational quantum algorithms.

\subsection{Robustness, Noise Scaling, and Ablation Studies}
\begin{figure}
	\includegraphics[width=\linewidth]{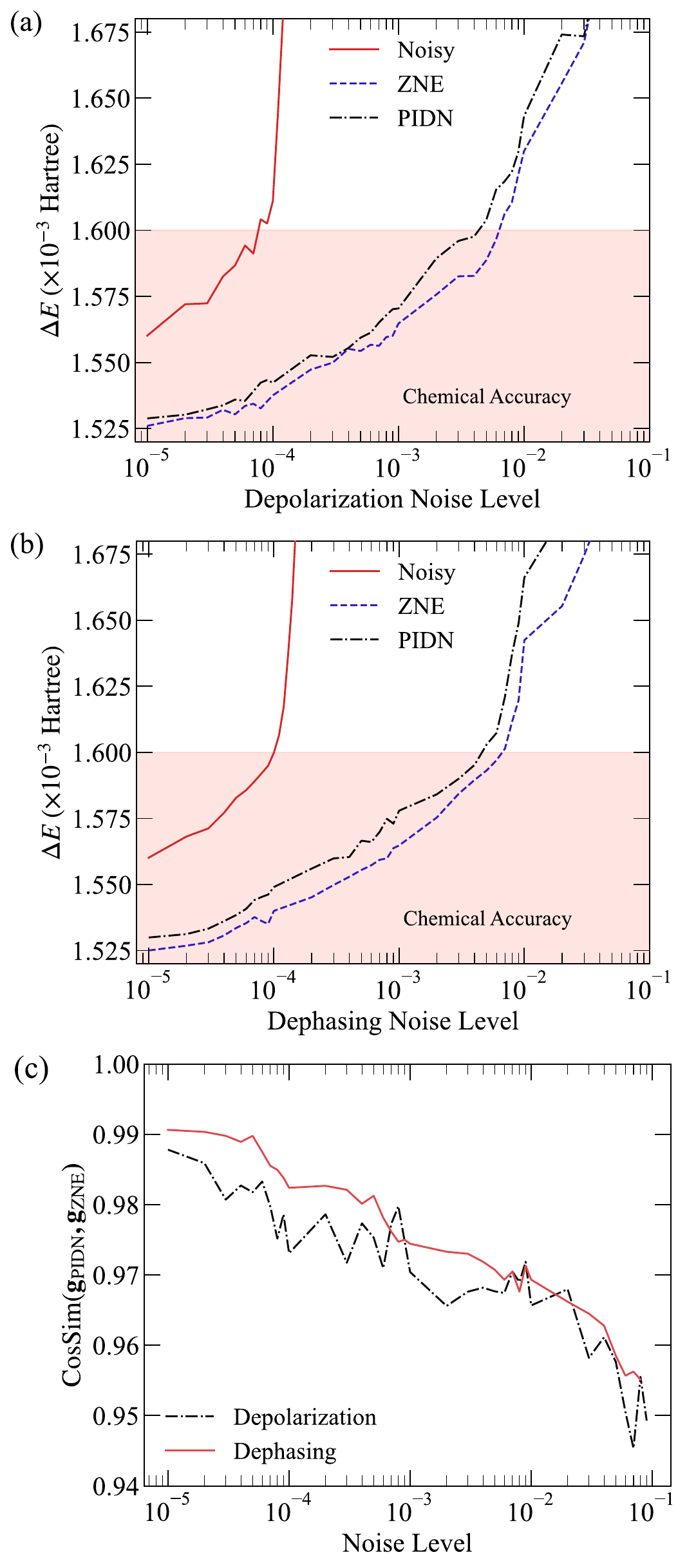}
	\caption{Final ground-state energy deviation for VQE on BeH$_2$ as a function of depolarizing and dephasing noise strength in (a) and (b)
(c) Average cosine similarity between PIDN-predicted gradients and ZNE gradients over the full optimization trajectory across noise levels.}
	\label{fig:Noise_scaling}
\end{figure}

In this subsection, we investigate two fundamental questions: (i) Under what noise regimes does PIDN-assisted optimization fail? and (ii) Is the physics-informed loss essential to its performance?

We first evaluate the performance of noisy VQE, ZNE-assisted VQE, and PIDN-assisted VQE for BeH$_2$ under depolarizing and dephasing noise levels ranging from $10^{-5}-10^{-1}$. Performance is measured by the final ground-state energy deviation relative to the ground-state energy calculated from full configuration interaction, and failure is declared when chemical accuracy is exceeded.

We observe the following regimes in Fig.~\ref{fig:Noise_scaling}(a) and (b): noisy VQE exceeds chemical accuracy at noise levels on the order of $10^{-4}$. In contrast, ZNE-assisted VQE maintains chemical accuracy up to noise levels around $10^{-2}$. PIDN-assisted VQE exhibits very similar behavior to ZNE, likewise failing to preserve chemical accuracy beyond approximately $10^{-2}$.

\begin{figure}
	\includegraphics[width=\linewidth]{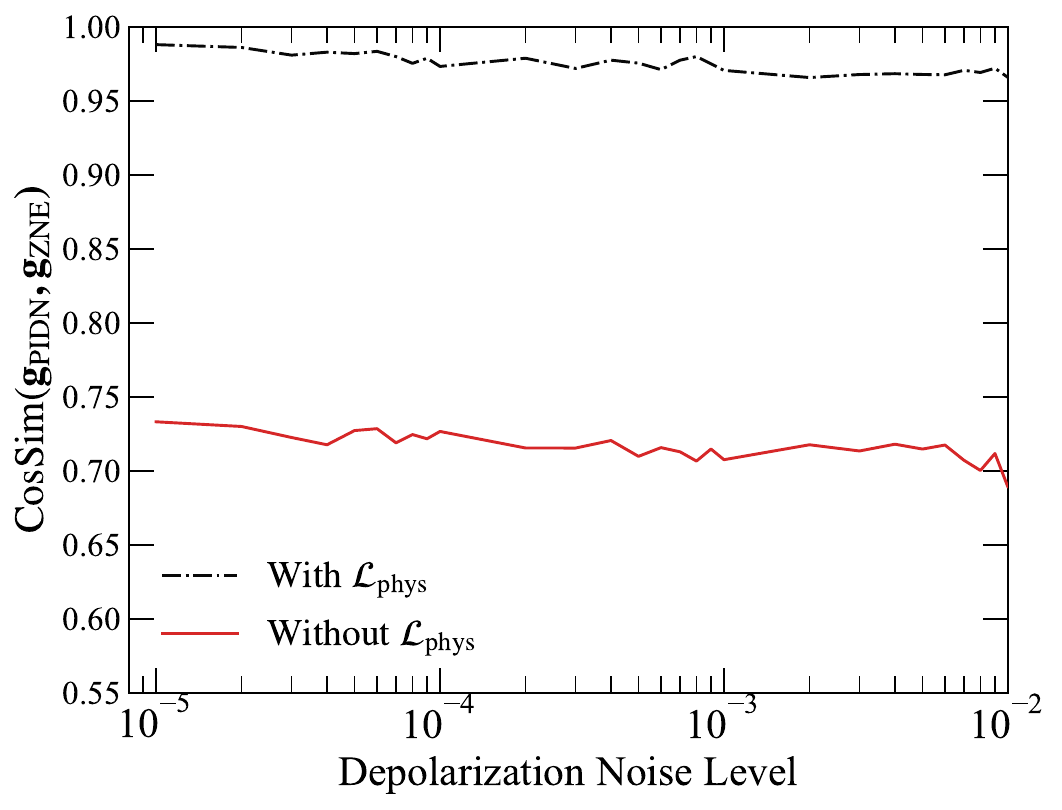}
	\caption{Average cosine similarity between PIDN-predicted gradients and ZNE gradients over the full optimization trajectory across noise levels, with models trained with and without the physics-informed loss.}
	\label{fig:phys_ab}
\end{figure}
Importantly, the performance curve of PIDN closely tracks that of ZNE across the entire noise range, consistently remaining slightly above ZNE's energy deviation. This behavior is expected, as PIDN is trained using ZNE-generated reference data and therefore approximates, but does not surpass ZNE performance.

A key observation is that the noise level at which PIDN fails is very close to where ZNE fails. This raises a critical question: Does PIDN fail because it cannot accurately reproduce ZNE dynamics under high noise, or does it fail because ZNE itself becomes unreliable?

To answer this question, we compute the average cosine similarity between PIDN-predicted gradients and ZNE gradients throughout training across varying noise levels. We find that the cosine similarity remains consistently above 0.95 across all noise regimes as shown in Fig.~\ref{fig:Noise_scaling}(c). This result indicates that PIDN continues to faithfully reproduce ZNE gradient directions even at high noise levels. Also, a mild decreasing trend is observed as noise strength increases, which can be attributed to the increased statistical variance in ZNE gradients at higher noise strengths since extrapolation becomes more sensitive to shot noise and extrapolation model mismatch.

This demonstrates that PIDN does not independently break down as a surrogate model. Instead, it faithfully tracks ZNE dynamics even when those dynamics themselves become less optimal. We therefore conclude that: PIDN fails when ZNE fails, not because it loses surrogate fidelity, but because it inherits the fundamental limitations of ZNE under strong noise.

We next investigate whether the physics-informed loss term is essential to PIDN's performance by comparing two training variants: Full PIDN with both supervised data loss and physics loss and ablated model with supervised data loss only.

We measure the cosine similarity between predicted and ZNE gradients under different noise levels as shown in Fig.~\ref{fig:phys_ab}, and find that with the physics loss, the cosine similarity is around $0.95$, whereas without the physics loss, it decreases to around $0.75$.

This substantial degradation indicates that the physics loss plays a critical role in constraining the surrogate to respect the underlying optimization trajectory. Without it, the network fails to preserve the correct vector-field geometry along the trajectory, leading to distorted gradient directions. Since optimization dynamics depend primarily on descent direction, not merely function approximation accuracy, this loss component is essential for maintaining stable training trajectories.

\begin{table}[t]
\centering
\begin{tabular}{ccc}
\hline
Graph / Problem & AR With $\mathcal{L}_\mathrm{phys}$ & AR Without $\mathcal{L}_\mathrm{phys}$ \\
\hline
TFIM            & 0.95              & 0.80                 \\

SK-model        & 0.92              & 0.78                 \\

3-regular       & 0.94              & 0.71                 \\
\hline
\end{tabular}
\caption{Impact of physics-informed loss on QAOA performance.
Final AR for QAOA on different problem instances, evaluated using ZNE under the final optimized parameters of PIDN with and without $\mathcal{L}_\mathrm{phys}$.}
\label{tab:qaoaphysics}
\end{table}

\begin{table}[b]
\centering
\begin{tabular}{ccc}
\hline
Molecule &$\Delta E$ with $\mathcal{L}_\mathrm{phys}$ & $\Delta E$ without $\mathcal{L}_\mathrm{phys}$ \\
\hline
LiH      & 0.85             & 28                \\

BeH$_2$  & 0.73             & 40                \\

H$_2$O   & 1.24             & 160                \\
\hline
\end{tabular}
\caption{Impact of physics-informed loss on VQE performance.
Final energy deviation $\Delta E$ (in mHa) for VQE on different molecules, evaluated using ZNE under the final optimized parameters of PIDN with and without $\mathcal{L}_\mathrm{phys}$.}
\label{tab:vqephysics}
\end{table}

To further quantify the impact of the physics-informed loss, we evaluate the final performance of PIDN-assisted optimization with and without $\mathcal{L}_{\mathrm{phys}}$. For QAOA, we measure the final AR using the ZNE-evaluated cost function under the parameters obtained by PIDN with and without $\mathcal{L}_{\mathrm{phys}}$ across three different problem instances: TFIM, SK model, and 3-regular graphs. As shown in TABLE~\ref{tab:qaoaphysics}, the inclusion of the physics loss consistently improves AR by a factor of $1.18 \sim 1.32$, confirming its positive effect. Similarly, for VQE tasks on molecular systems (LiH, BeH$_2$ and H$_2$O), we evaluate the final energy deviation relative to the noiseless reference using ZNE. Table~\ref{tab:vqephysics} shows that the physics loss reduces the energy deviation by  a factor of $33\sim129$ across all molecules, highlighting its importance in preserving accurate gradient directions and guiding the optimizer toward lower-energy configurations.

\begin{figure}
	\includegraphics[width=\linewidth]{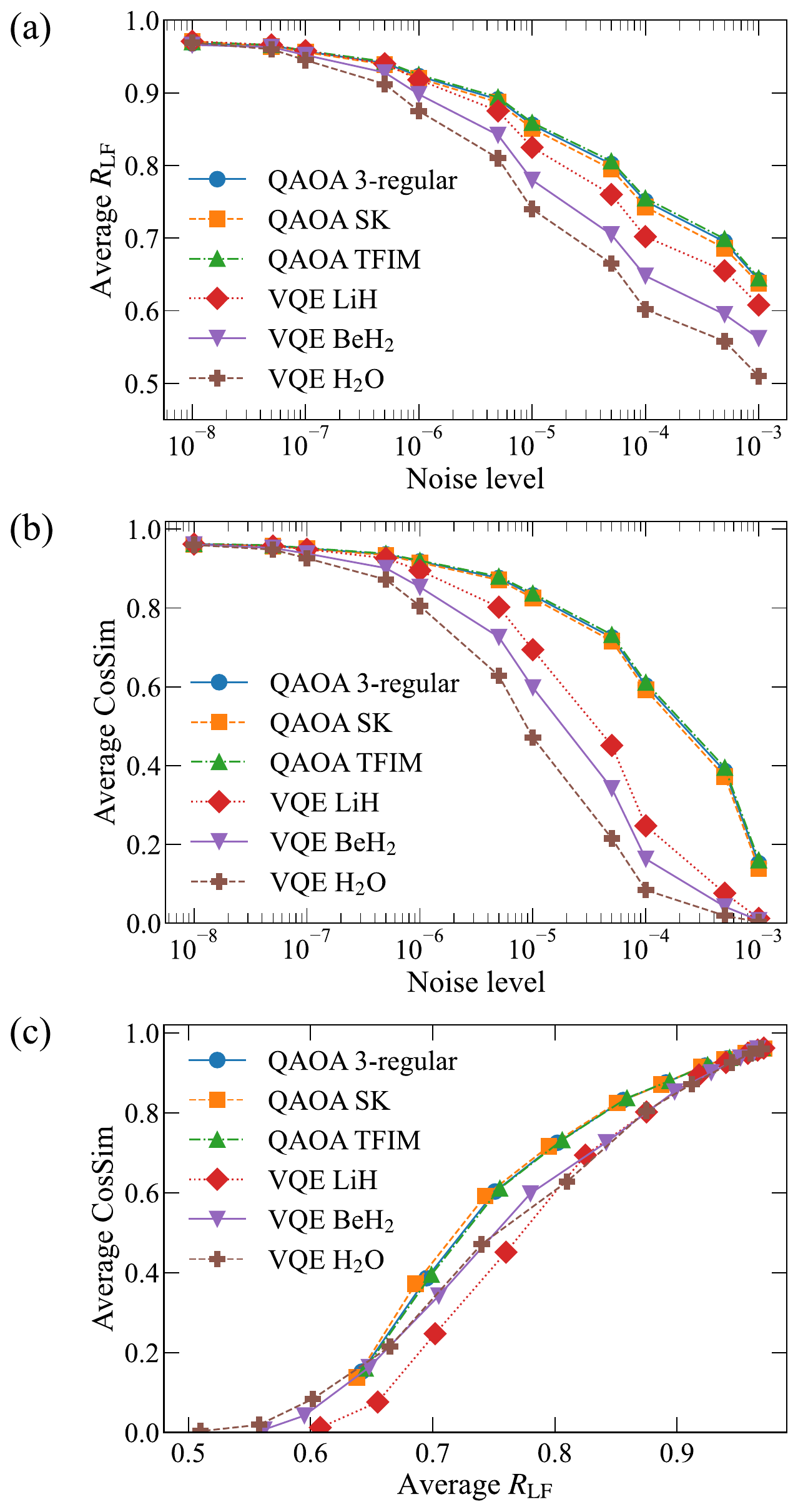}
	\caption{Dependence of PIDN trajectory-tracking accuracy on the spectral smoothness of the noisy loss landscape. (a) Average low-frequency power ratio $R_{\mathrm{LF}}$ as a function of noise strength for different QAOA and VQE tasks. (b) Average cosine similarity between PIDN-predicted update directions and the corresponding reference noisy gradients as a function of noise strength. (c) Average cosine similarity plotted against $R_{\mathrm{LF}}$. A clear positive correlation is observed across all tasks.}
	\label{fig:roughness_tracking}
\end{figure}

To further understand why PIDN is able to track the optimization direction using data from the early stage of training, we examine the role of the underlying loss-landscape geometry. 

The central hypothesis is that PIDN can effectively track the optimization dynamics when the local geometry of the loss landscape varies smoothly along the trajectory. In this case, the information contained in the early-stage optimization path, together with the currently measured noisy expectation value, is sufficient to infer a ZNE-like update direction in the nearby region of parameter space. By contrast, if the loss landscape changes rapidly, or equivalently if it contains substantial high-frequency structure, then the local optimization flow becomes much harder to extrapolate from the earlier trajectory. Under such conditions, even small displacements in parameter space can lead to substantial changes in gradient direction, which reduces the ability of PIDN to maintain accurate trajectory-level tracking.

To test this idea, we perform a controlled study in which no error-mitigation method is applied and only noisy expectation values are used. The results are summarized in Tables~\ref{tab:rlf_noise_scaling} and \ref{tab:cossim_noise_scaling}, and are visualized in Fig.~\ref{fig:roughness_tracking}, where panels (a) and (b) show the dependence of $R_{\mathrm{LF}}$ (see Eq.~\eqref{eq:RLF}) and cosine similarity on noise strength, respectively, and panel (c) directly compares the average cosine similarity with $R_{\mathrm{LF}}$.

\begin{table*}[b]
\centering
\renewcommand{\arraystretch}{1.15}
\begin{tabular}{c|ccc|ccc}
\hline
\multirow{2}{*}{Noise level} 
& \multicolumn{3}{c|}{QAOA} 
& \multicolumn{3}{c}{VQE} \\
\cline{2-7}
& 3-regular & SK model & TFIM & LiH & BeH$_2$ & H$_2$O \\
\hline
$1\times10^{-8}$  & 0.968 & 0.971 & 0.970 & 0.971 & 0.966 & 0.969 \\
$5\times10^{-8}$  & 0.965 & 0.964 & 0.966 & 0.966 & 0.963 & 0.960 \\
$1\times10^{-7}$  & 0.957 & 0.956 & 0.958 & 0.958 & 0.952 & 0.945 \\
$5\times10^{-7}$  & 0.941 & 0.939 & 0.943 & 0.940 & 0.928 & 0.912 \\
$1\times10^{-6}$  & 0.923 & 0.920 & 0.925 & 0.918 & 0.898 & 0.875 \\
$5\times10^{-6}$  & 0.891 & 0.887 & 0.894 & 0.875 & 0.842 & 0.810 \\
$1\times10^{-5}$  & 0.856 & 0.851 & 0.859 & 0.825 & 0.780 & 0.740 \\
$5\times10^{-5}$  & 0.802 & 0.795 & 0.806 & 0.760 & 0.705 & 0.665 \\
$1\times10^{-4}$  & 0.751 & 0.743 & 0.755 & 0.702 & 0.648 & 0.602 \\
$5\times10^{-4}$  & 0.695 & 0.686 & 0.699 & 0.655 & 0.595 & 0.558 \\
$1\times10^{-3}$  & 0.642 & 0.638 & 0.645 & 0.608 & 0.562 & 0.510 \\
\hline
\end{tabular}
\caption{Average low-frequency power ratio $R_{\mathrm{LF}}$ of the noisy loss landscape as a function of noise strength for different VQA tasks. For each case, all PQC parameters are randomly fixed except for eight selected parameters, and the reported value is averaged over multiple random parameter fixings and different choices of the eight active parameters. Larger $R_{\mathrm{LF}}$ indicates that the landscape is more strongly dominated by smooth low-frequency structure, whereas smaller values reflect increasing high-frequency distortion.}
\label{tab:rlf_noise_scaling}
\end{table*}

\begin{table*}[ht]
\centering
\renewcommand{\arraystretch}{1.15}
\begin{tabular}{c|ccc|ccc}
\hline
\multirow{2}{*}{Noise level} 
& \multicolumn{3}{c|}{QAOA} 
& \multicolumn{3}{c}{VQE} \\
\cline{2-7}
& 3-regular & SK model & TFIM & LiH & BeH$_2$ & H$_2$O \\
\hline
$1\times10^{-8}$  & 0.962 & 0.961 & 0.963 & 0.962 & 0.961 & 0.960 \\
$5\times10^{-8}$  & 0.958 & 0.957 & 0.959 & 0.957 & 0.952 & 0.948 \\
$1\times10^{-7}$  & 0.951 & 0.950 & 0.952 & 0.949 & 0.938 & 0.926 \\
$5\times10^{-7}$  & 0.936 & 0.934 & 0.938 & 0.926 & 0.901 & 0.872 \\
$1\times10^{-6}$  & 0.918 & 0.915 & 0.920 & 0.895 & 0.853 & 0.805 \\
$5\times10^{-6}$  & 0.876 & 0.871 & 0.880 & 0.802 & 0.726 & 0.628 \\
$1\times10^{-5}$  & 0.832 & 0.825 & 0.837 & 0.694 & 0.598 & 0.471 \\
$5\times10^{-5}$  & 0.725 & 0.716 & 0.732 & 0.451 & 0.342 & 0.215 \\
$1\times10^{-4}$  & 0.603 & 0.592 & 0.611 & 0.247 & 0.163 & 0.084 \\
$5\times10^{-4}$  & 0.386 & 0.372 & 0.395 & 0.076 & 0.042 & 0.019 \\
$1\times10^{-3}$  & 0.152 & 0.138 & 0.160 & 0.012 & 0.007 & 0.003 \\
\hline
\end{tabular}
\caption{Average cosine similarity between PIDN-predicted update directions and reference noisy gradients at different noise levels for different VQA tasks. Here, PIDN is trained using the first few iterations of the noisy optimization trajectory and evaluated on the subsequent trajectory-continuation stage.}
\label{tab:cossim_noise_scaling}
\end{table*}

Starting from the weak-noise regime at $10^{-8}$ and gradually increasing the noise strength to $10^{-3}$, we examine how the geometry of the noisy loss landscape changes and how this affects the trajectory-tracking ability of PIDN. As shown in Fig.~\ref{fig:roughness_tracking}(a), the low-frequency power ratio $R_{\mathrm{LF}}$ decreases monotonically with increasing noise for all tasks, indicating that the smooth large-scale structure of the landscape is progressively weakened while higher-frequency distortion becomes more significant. For QAOA, this reduction is relatively gradual: in the 3-regular case, $R_{\mathrm{LF}}$ decreases from $0.968$ at $10^{-8}$ to $0.642$ at $10^{-3}$. Similar behavior is observed for the SK model and TFIM, where $R_{\mathrm{LF}}$ decreases from $0.971$ to $0.645$, over the same noise range. For VQE, the degradation is slightly more pronounced: in LiH, $R_{\mathrm{LF}}$ decreases from $0.971$ at $10^{-8}$ to $0.608$ at $10^{-3}$; in BeH$_2$, the corresponding values are $0.966$ and $0.562$; and in H$_2$O they are $0.969$ and $0.510$. These results show that increasing noise progressively transfers spectral weight away from the low-frequency sector and renders the effective landscape increasingly rough and irregular.

We then train PIDN on the first few iterations of the noisy optimization trajectory and evaluate its ability to continue the subsequent updates. The agreement between PIDN-predicted update directions and the corresponding noisy reference gradients is quantified by the cosine similarity defined in Eq.~\eqref{eq:cossim}. As shown in Fig.~\ref{fig:roughness_tracking}(b), the average cosine similarity decreases systematically as the noise strength increases, mirroring the behavior of $R_{\mathrm{LF}}$. For QAOA, PIDN retains strong directional agreement throughout the weak- and intermediate-noise regimes: in the 3-regular case, the cosine similarity is $0.962$ at $10^{-8}$, remains as high as $0.918$ at $10^{-6}$, then decreases to $0.603$ at $10^{-4}$ and $0.152$ at $10^{-3}$. Similar trends are obtained for the SK model, where the cosine similarity decreases from $0.961$ to $0.915$, $0.592$, and $0.138$, and for TFIM, where it decreases from $0.963$ to $0.920$, $0.611$, and $0.160$. For VQE, the deterioration is substantially faster. In LiH, the cosine similarity drops from $0.962$ at $10^{-8}$ to $0.802$ at $10^{-6}$, then to $0.247$ at $10^{-4}$ and only $0.012$ at $10^{-3}$. For BeH$_2$, it decreases from $0.961$ to $0.853$, $0.163$, and $0.007$, while for H$_2$O it decreases from $0.960$ to $0.805$, $0.084$, and $0.003$. The connection between these two quantities is shown most directly in Fig.~\ref{fig:roughness_tracking}(c), where a clear positive correlation is observed between the average cosine similarity and $R_{\mathrm{LF}}$ across all tasks. These results indicate that although PIDN can reliably track the local optimization flow when the landscape retains substantial low-frequency structure, its trajectory-continuation accuracy deteriorates once noise-induced high-frequency distortion becomes dominant. These results support the interpretation that PIDN performs best when the effective loss landscape remains spectrally smooth, as in the case of ZNE-mitigated VQAs, and that the gradual loss of low-frequency structure directly limits the ability of early-stage trajectory information to determine later optimization directions.

\section{Conclusion and Outlook} \label{sec:conclusion}
In this work, we have presented a PIDN framework to accelerate noisy VQAs, including QAOA and VQE, using ZNE as a reference. We formulated the optimization process as a dynamical trajectory in parameter space, where the network learns to predict both denoised expectation values and the next-step parameters. Our results demonstrate that PIDN can faithfully emulate the gradient dynamics of ZNE, effectively reducing the need for costly repeated circuit executions while preserving convergence toward high-quality solutions. Through landscape reconstruction and trajectory analysis, we have shown that PIDN captures the essential features of noiseless or ZNE-assisted surfaces, including both global trends and local curvature, thereby guiding the optimizer efficiently even under significant hardware noise.

We further quantified the efficiency gains of PIDN across a range of VQA tasks. For QAOA on 3-regular graphs, SK models, and TFIM instances, PIDN achieved approximation ratios comparable to ZNE-assisted optimization while reducing the total number of circuit executions by approximately a factor of five. For VQE tasks on LiH, BeH$_2$, and H$_2$O molecules, similar speedups in the range of four- to six-fold were observed while maintaining chemical accuracy. These results underscore the generality and robustness of the approach across both combinatorial optimization and molecular simulation problems. By systematically analyzing shot efficiency, evaluation trends, and final performance metrics, we provide strong evidence that PIDN offers a scalable strategy to mitigate noise-induced barren plateaus without compromising solution quality.

Our robustness, ablation, and landscape-geometry analyses clarify the mechanisms underlying PIDN's performance. By examining energy deviations and gradient cosine similarity across different noise levels, we find that PIDN fails only when ZNE itself becomes unreliable, indicating that its limitations are inherited from the reference mitigation method rather than intrinsic to the surrogate model. The ablation study further shows that the physics-informed loss is essential for preserving high directional alignment between the predicted and ZNE gradients, leading to clear improvements in both final energy and approximation ratio, highlighting the importance of incorporating physical priors into surrogate models to preserve optimization dynamics in the NISQ regime. In addition, our controlled analysis of noisy optimization without error mitigation shows that PIDN tracks the optimization flow most accurately when the effective loss landscape retains strong low-frequency structure. As the noise level increases, the low-frequency power ratio decreases, high-frequency distortion becomes more pronounced, and the cosine similarity between PIDN-predicted and reference gradients deteriorates systematically. These results show that the success of PIDN relies on the persistence of sufficiently smooth landscape geometry from accurate denoising.

The PIDN framework offers several avenues for further research. Extending the method to multi-parameter extrapolation schemes, adaptive online retraining strategies, or more complex ansatz may further reduce resource overheads while improving robustness. Additionally, combining PIDN with other noise-mitigation or quantum error suppression techniques could enable near-term quantum devices to tackle larger combinatorial or chemical problems with reduced computational cost. Beyond VQA, the general principle of physics-informed surrogate modeling could find applications in hybrid quantum-classical control, quantum metrology, and variational simulations of time-dependent Hamiltonians, making PIDN a useful tool for the broader quantum computing community.

\section*{ACKNOWLEDGEMENTS}
This work is supported by the National Natural Science Foundation of China (Grant No. 12474489), Shenzhen Fundamental Research Program (Grant No. JCYJ20240813153139050), and the Guangdong Provincial Quantum Science Strategic Initiative (Grant No. GDZX2203001, GDZX2403001).

\section*{DATA AVAILABILITY}
The data that support the findings of this article are openly available \cite{data}.

\end{document}